\documentclass[12pt]{article}

\usepackage{graphicx}
\usepackage{amsmath}
\usepackage{amssymb}
\usepackage{amsthm}
\numberwithin{equation}{section}

\newcommand{\cO}{{\mathcal O}}
\newcommand{\cT}{{\mathcal T}}
\newcommand{\cQ}{{\mathcal Q}}
\newcommand{\sC}{{C\!\!\!\!/\,}}
\newcommand{\fJ}{{\mathfrak J}}
\newcommand{\fR}{{\mathfrak R}}
\newcommand{\fL}{{\mathfrak L}}
\newcommand{\fC}{{\mathfrak C}}
\newcommand{\fB}{{\mathfrak B}}
\newcommand{\fF}{{\mathfrak F}}
\newcommand{\sfR}{{\mathfrak R}\!\!\!\!/\,}
\newcommand{\sfL}{{\mathfrak L}\!\!\!/}
\newcommand{\sfC}{{\mathfrak C}\!\!\!/}
\newcommand{\Tr}{{\rm Tr}}
\newcommand{\bs}[1]{\boldsymbol{#1}}
\newcommand{\alg}[1]{\mathfrak{#1}}

\begin{document}

\baselineskip=16pt plus 0.2pt minus 0.1pt

\begin{titlepage}
\title{
\vspace{1cm}
{\Large\bf Serre Relation and Higher Grade Generators\\
of the AdS/CFT Yangian Symmetry}
}
\vskip20mm
\author{
{\sc Takuya Matsumoto}\thanks
{{\tt m05044c@math.nagoya-u.ac.jp}}\;
\quad and \quad
{\sc Sanefumi Moriyama}\thanks
{{\tt moriyama@math.nagoya-u.ac.jp}}\\[15pt]
{\it Graduate School of Mathematics, Nagoya University,} \\
{\it Nagoya 464-8602, Japan}
}
\date{\normalsize February, 2009}
\maketitle
\thispagestyle{empty}
\begin{abstract}
\normalsize
It was shown that the spin chain model coming from AdS/CFT
correspondence satisfies the Yangian symmetry if we assume
evaluation representation, though so far there is no explicit proof
that the evaluation representation satisfies the Serre relation, which
is one of the defining equations of the Yangian algebra imposing
constraints on the whole algebraic structure.
We prove completely that the evaluation representation adopted in the
model satisfies the Serre relation by introducing a three-dimensional
gamma matrix.
After studying the Serre relation, we proceed to the whole Yangian
algebraic structure, where we find that the conventional construction
of higher grade generators is singular and we propose an alternative
construction.
In the discussion of the higher grade generators, a great
simplification for the proof of the Serre relation is found. 
Using this expression, we further show that the proof is lifted to the
exceptional superalgebra, which is a non-degenerate deformation of the
original superalgebra.
\end{abstract}
\end{titlepage}

\section{Introduction}

It is no doubt that Yang-Mills theory plays important roles in
understanding the modern particle physics.
Especially, the maximally super Yang-Mills theory enjoys an
interesting theoretical property.
Since the maximally super Yang-Mills theory is conjectured to be dual
to string theory on $AdS_5\times S^5$ with RR flux background
\cite{M}, it is expected that the super Yang-Mills theory can be used
to describe string theory as various matrix models describe various
string theories.

In this context, it is exciting to find \cite{MZ} that the one-loop
anomalous dimension of Yang-Mills theory is mapped explicitly to the
Hamiltonian of an integrable spin chain model, because infinite
generators of the integrable model may play a central role in studying
string theory on the curved spacetime\footnote{More directly, the
study of integrability in the string worldsheet theory was initiated
in \cite{MSW,BPR,HY}.
Due to the difficulty of the $\kappa$-symmetry, here we would like to
explore the integrability from the Yang-Mills side.},
as the Virasoro generators play on the flat spacetime.
Therefore, it is aspiring to understand the whole integrable algebraic
structure with infinite generators completely.

The analysis is pushed further to the all-loop level.
Fixing one direction as the vacuum, the original dynamic spin chain of 
the superconformal symmetry $\alg{psu}(2,2|4)$ is broken down to two
copies of undynamic ones of $\alg{psu}(2|2)$ with some common central
charges \cite{Bthesis,Bundyn}.
Especially, an off-shell formalism with the centrally extended
superalgebra $\alg{psu}(2|2)\ltimes{\mathbb R}^3$ was proposed in
\cite{BRmat}.
Then, it is surprising to find that the R-matrix\footnote{The R-matrix
is related to the scattering matrix in the context of physics.
Usually, the R-matrix is determined from both the Lie algebra and the
grade-1 Yangian algebra.
We believe that the reason that the R-matrix is determined only from
the Lie algebra is because the off-shell formalism with three central
charges contains the non-local information which is usually carried by
the Yangian algebra.}, 
which is determined uniquely (up to an overall factor) by requiring
this Lie algebraic symmetry $\alg{psu}(2|2)\ltimes{\mathbb R}^3$,
enjoys the Yang-Baxter equation, which implies the integrability.
Besides, it was also shown \cite{BYangian} that the R-matrix has the
grade-1 Yangian symmetry if we assume the evaluation representation
for the Yangian algebra.
Later, the Yangian algebra with the evaluation representation also
turns out to be crucial in determining the R-matrix of bound states
\cite{bound}.

The Yangian algebra consists of infinitely graded generators
$\fJ^A_n$.
(See \cite{D} for the original work and \cite{MK} for a review.)
The grade-0 generator $\fJ^A_0=\fJ^A$ is that of the standard Lie
algebra $[\fJ^A,\fJ^B]=\fJ^Cf_C{}^{AB}$ (with the superscript on the
left and the subscript on the right, as the canonical position of the
spinor indices), satisfying the Jacobi identity,
\begin{align}
[\fJ^A,[\fJ^B,\fJ^C]]+(-)^{|A|(|B|+|C|)}[\fJ^B,[\fJ^C,\fJ^A]]
+(-)^{|C|(|A|+|B|)}[\fJ^C,[\fJ^A,\fJ^B]]=0.
\label{jacobi}
\end{align}
Note that we consider the superalgebra, where the commutator and the
anticommutator (which will appear later) denote the generalized ones
$[\fJ^A,\fJ^B]=\fJ^A\fJ^B-(-)^{|A||B|}\fJ^B\fJ^A$,
$\{\fJ^A,\fJ^B\}=\fJ^A\fJ^B+(-)^{|A||B|}\fJ^B\fJ^A$, with $|A|=0$
$(|A|=1)$ for the bosonic (fermionic) generator $\fJ^A$.

The grade-1 generator $\fJ^A_1=\widehat\fJ^A$, with the commutation
relations with the Lie algebra generator
$[\widehat\fJ^A,\fJ^B]=[\fJ^A,\widehat\fJ^B]=\widehat\fJ^Cf_C{}^{AB}$,
satisfies further the Serre relation,
\begin{align}
&[\widehat\fJ^A,[\widehat\fJ^B,\fJ^C]]
+(-)^{|A|(|B|+|C|)}[\widehat\fJ^B,[\widehat\fJ^C,\fJ^A]]
+(-)^{|C|(|A|+|B|)}[\widehat\fJ^C,[\widehat\fJ^A,\fJ^B]]\nonumber\\
&\quad=-\frac{\hbar^2}{24}(-)^{|I|(|B|+|J|+|C|+|K|)+|J|(|C|+|K|)}
\{\fJ^L,\fJ^M,\fJ^N\}f_L{}^{AI}f_M{}^{BJ}f_N{}^{CK}f_{IJK},
\label{serre}
\end{align}
where the generalized totally symmetrized product is
\begin{align}
\{\fJ^L,\fJ^M,\fJ^N\}
=\{\fJ^L,\fJ^M\}\fJ^N+(-)^{|L|(|M|+|N|)}\{\fJ^M,\fJ^N\}\fJ^L
+(-)^{|N|(|L|+|M|)}\{\fJ^N,\fJ^L\}\fJ^M.
\end{align}
Here the complicated-looking sign
$(-)^{|I|(|B|+|J|+|C|+|K|)+|J|(|C|+|K|)}$ in \eqref{serre} can be
simply interpreted as the Grassmannian charge which appears if we
exchange the order of the indices to contract the subscripts $I,J,K$
in $f_{IJK}$ with the same superscripts in
$\fJ^Lf_L{}^{AI}\fJ^Mf_M{}^{BJ}\fJ^Nf_N{}^{CK}$ in the canonical
position.
Serre relation in the simple Lie algebra imposes constraints on the
construction of full root system from the simple roots.
Here in the Yangian algebra, the Serre relation \eqref{serre} puts
constraints on the construction of higher grade generators.
Therefore, it is obvious that the Serre relation is important in
studying the whole integrable algebraic structure with the higher
grade generators.

Some representations of the Lie algebra $\alg{g}$ admit the evaluation
representations for the corresponding Yangian algebra $Y(\alg{g})$,
which require the representation of the higher grade generators to
relate to that of the Lie algebra by $\fJ^A_n\simeq(iu)^n\fJ^A$ with
the spectral parameter $u$.
The physical meaning of the evaluation representation is not very
clear, but since the evaluation representation takes a similar
appearance as the affine Lie algebra of conformal field theory and
since the classical limit of the R-matrix with the Yangian symmetry
takes the canonical form
\begin{align}
r_{12}=\frac{\cT^\alg{g}_{12}}{i(u_1-u_2)}
=\sum_{n=0}^\infty\fJ^A_{-n-1}\otimes\fJ_{An},
\end{align}
with the two-site Casimir operator
$\cT^\alg{g}_{12}=\fJ^A\otimes\fJ_A=\fJ^A\otimes\fJ^Bg_{BA}$, which is
reminiscent of the KZ equation of the WZW model, it is natural to
expect that the evaluation representation has a close relation with
the classical limit of the string worldsheet theory.
The classical limit of the $\alg{psu}(2|2)\ltimes{\mathbb R}^3$ spin
chain model was investigated in \cite{classical,MMT,BS,MM}.

Although it was found that the R-matrix has the grade-1 Yangian
coproduct symmetry if we adopt the evaluation representation, it has
not been shown explicitly so far that the evaluation representation is
compatible with the Serre relation\footnote{The relation between the
current Drinfeld first realization and an alternative Drinfeld second
realization was investigated in \cite{ST}, where the evaluation
representation in the second realization was also studied.
It was shown that two different evaluation parameters have to be
introduced.
In our following analysis in the first realization, we do not see any
signs of the appearance of two evaluation parameters.
We would like to see the relation between these two facts and the
possibility of avoiding the situation with two parameters even in the
second realization.}.
One of the main subjects in this paper is to study the compatibility.
Namely, since, in the evaluation representation
$\widehat{\fJ}^A\simeq iu\fJ^A$, the left-hand-side of the Serre
relation \eqref{serre} reduces to that of the Jacobi identity
\eqref{jacobi}, the right-hand-side of the Serre relation
\eqref{serre} has to vanish when acting on any states.
We will prove it extensively in the following sections.

One reason that this computation has not been done before is because
of the degeneracy of the Killing form of the superalgebra
$\alg{psu}(2|2)\ltimes{\mathbb R}^3$.
In studying the Serre relation, we have to raise and lower the indices 
frequently and it is impossible to proceed if the Killing form is
degenerate.
In \cite{BRmat,MM}, it was shown that the algebra is a special limit
($\varepsilon\to 0$) of the exceptional superalgebra
$\alg{d}(2,1;\varepsilon)$ with a parameter $\varepsilon$, whose
Killing form is non-degenerate and, therefore, many algebraic
properties can be studied in this $\varepsilon$-deformation.
Motivated by its non-degeneracy, we also constructed an
infinite-dimensional representation of the exceptional superalgebra
$\alg{d}(2,1;\varepsilon)$ in \cite{MM}, which is a natural lift of
the fundamental representation of the superalgebra
$\alg{psu}(2|2)\ltimes{\mathbb R}^3$.

Another difficulty is due to the complexity of the structure
constants.
In order to study the Serre relation, let us read off the Killing form
$g_{AB}$ from the two-site Casimir operator\footnote{We have omitted
the braiding factors appearing in the tensor product for notational
simplicity, which are actually necessary in the off-shell formalism
\cite{BRmat,braiding}.},
\begin{align}
\cT^{\alg{d}}_{12}=-\alpha\fR^a{}_b\otimes\fR^b{}_a
-\beta\fL^\alpha{}_\beta\otimes\fL^\beta{}_\alpha
-\gamma\fC^\alg{a}{}_\alg{b}\otimes\fC^\alg{b}{}_\alg{a}
-\epsilon_{ab}\epsilon_{\alpha\beta}\epsilon_{\alg{a}\alg{b}}
\fF^{a\alpha\alg{a}}\otimes\fF^{b\beta\alg{b}},
\label{casimir}
\end{align}
and spell out the non-zero structure constant from the commutation
relations
\begin{align}
f_{\fR^a{}_{a'}\fR^b{}_{b'}\fR^c{}_{c'}}
&=\alpha^2(\delta_a^{b'}\delta_b^{c'}\delta_c^{a'}
-\delta_a^{c'}\delta_b^{a'}\delta_c^{b'}),
&
f_{\fF^{a\alpha\alg{a}}\fF^{b\beta\alg{b}}\fR^c{}_{c'}}
&=\frac{\alpha}{2}
(\epsilon_{ac}\delta^{c'}_b+\epsilon_{bc}\delta^{c'}_a)
\epsilon_{\alpha\beta}\epsilon_{\alg{a}\alg{b}},
\nonumber\\
f_{\fL^\alpha{}_{\alpha'}\fL^\beta{}_{\beta'}\fL^\gamma{}_{\gamma'}}
&=\beta^2
(\delta_\alpha^{\beta'}\delta_\beta^{\gamma'}\delta_\gamma^{\alpha'}
-\delta_\alpha^{\gamma'}\delta_\beta^{\alpha'}\delta_\gamma^{\beta'}),
&
f_{\fF^{a\alpha\alg{a}}\fF^{b\beta\alg{b}}\fL^\gamma{}_{\gamma'}}
&=\frac{\beta}{2}\epsilon_{ab}
(\epsilon_{\alpha\gamma}\delta^{\gamma'}_\beta
+\epsilon_{\beta\gamma}\delta^{\gamma'}_\alpha)
\epsilon_{\alg{a}\alg{b}},
\nonumber\\
f_{\fC^\alg{a}{}_{\alg{a}'}\fC^\alg{b}{}_{\alg{b}'}
\fC^\alg{c}{}_{\alg{c}'}}
&=\gamma^2(\delta_\alg{a}^{\alg{b}'}\delta_\alg{b}^{\alg{c}'}
\delta_\alg{c}^{\alg{a}'}
-\delta_\alg{a}^{\alg{c}'}\delta_\alg{b}^{\alg{a}'}
\delta_\alg{c}^{\alg{b}'}),
&
f_{\fF^{a\alpha\alg{a}}\fF^{b\beta\alg{b}}\fC^\alg{c}{}_{\alg{c}'}}
&=\frac{\gamma}{2}\epsilon_{ab}\epsilon_{\alpha\beta}
(\epsilon_{\alg{a}\alg{c}}\delta^{\alg{c}'}_\alg{b}
+\epsilon_{\alg{b}\alg{c}}\delta^{\alg{c}'}_\alg{a}).
\label{strconst}
\end{align}
If we plug these structure constants into the Serre relation, we can 
easily get stranded because of the complexity of $2^4=16$ terms, which
comes from four structure constants with each consisting of two terms.

In this paper\footnote{According to \cite{DNW}, there is a standard
method to prove the Serre relation for the evaluation representation
of the Yangian algebra $Y(\alg{su}(n))$ using the so-called nesting
relation.
We do not adopt the method here because we are also interested in the
question whether the representation of the exceptional algebra
$\alg{d}(2,1;\varepsilon)$ constructed in \cite{MM} admit the
evaluation representation, where the method seems not to be
applicable.},
we shall regard $\alg{su}(2)$ as $\alg{so}(3)$ and define a
three-dimensional gamma matrix.
Then, all of the structure constants \eqref{strconst} can be rewritten
in a simple form in terms of the gamma matrix.
With lots of formulas of the gamma matrix, the computation can be done
without difficulty as that of scattering amplitudes of fermions.

After proving that the evaluation representation is compatible with
the Serre relation, we proceed to construct higher grade generators.
We find a novel subtle structure, suggesting that the canonical
expression of the higher grade generators has to be improved.
We propose a resolution to this subtlety.
Our resolution further implies a great simplification for the proof of
the compatibility.
Hence, subsequently, we come back to the non-degenerate deformation of
the exceptional superalgebra $\alg{d}(2,1;\varepsilon)$ and prove that
the compatibility of the evaluation representation is also lifted the
exceptional superalgebra $\alg{d}(2,1;\varepsilon)$.

In the next section, after recapitulating the algebra shortly, we
reformulate the algebra and the representation by introducing a
three-dimensional gamma matrix.
Then, in section 3, we prove that the evaluation representation is
compatible with the Serre relation.
After the proof, we proceed to the higher grade generators in section
4.
And in section 5 we present a proof of the compatibility for the case
of the exceptional superalgebra.
Finally, we conclude in section 6.
In appendix A, we show how the Serre relation relates to the
homomorphism of the coproduct.
The subsequent three appendices are devoted to various useful
formulas in computation.

\section{Gamma matrix formalism}
We shall introduce a three-dimensional gamma matrix and reformulate
the exceptional Lie superalgebra $\alg{d}(2,1;\varepsilon)$, its
$\varepsilon\to 0$ limit $\alg{psu}(2|2)\ltimes{\mathbb R}^3$ and 
their representations \cite{BRmat,MM} in terms of the gamma matrix.

\subsection{Algebra}
The generators of the exceptional Lie superalgebra
$\alg{d}(2,1;\varepsilon)$ consist of three orthogonal sets of
$\alg{su}(2)$ triplet bosonic generators $\fR^a{}_b$,
$\fL^\alpha{}_\beta$, $\fC^{\alg{a}}{}_{\alg{b}}$ (subject to the
traceless condition
$\fR^a{}_a=\fL^\alpha{}_\alpha=\fC^{\alg{a}}{}_{\alg{a}}=0$) and an
octet of fermionic generators $\fF^{a\alpha\alg{a}}$ where all indices
$a,\alpha,\alg{a}$ run over $1$ and $2$.
The non-trivial commutation relations are given as
\begin{align}
&[\fR^a{}_b,\fR^c{}_d]=\delta^c_b\fR^a{}_d-\delta^a_d\fR^c{}_b,\;
[\fL^\alpha{}_\beta,\fL^\gamma{}_\delta]
=\delta^\gamma_\beta\fL^\alpha{}_\delta
-\delta^\alpha_\delta\fL^\gamma{}_\beta,\;
[\fC^\alg{a}{}_\alg{b},\fC^\alg{c}{}_\alg{d}]
=\delta^\alg{c}_\alg{b}\fC^\alg{a}{}_\alg{d}
-\delta^\alg{a}_\alg{d}\fC^\alg{c}{}_\alg{b},\nonumber\\
&[\fR^a{}_b,\fF^{c\gamma\alg{c}}]
=\delta^c_b\fF^{a\gamma\alg{c}}
-\frac{1}{2}\delta^a_b\fF^{c\gamma\alg{c}},\;
[\fL^\alpha{}_\beta,\fF^{c\gamma\alg{c}}]
=\delta^\gamma_\beta\fF^{c\alpha\alg{c}}
-\frac{1}{2}\delta^\alpha_\beta\fF^{c\gamma\alg{c}},\;
[\fC^\alg{a}{}_\alg{b},\fF^{c\gamma\alg{c}}]
=\delta^\alg{c}_\alg{b}\fF^{c\gamma\alg{a}}
-\frac{1}{2}\delta^\alg{a}_\alg{b}\fF^{c\gamma\alg{c}},\nonumber\\
&[\fF^{a\alpha\alg{a}},\fF^{b\beta\alg{b}}]
=\alpha\epsilon^{ak}\epsilon^{\alpha\beta}\epsilon^{\alg{a}\alg{b}}
\fR^b{}_k
+\beta\epsilon^{ab}\epsilon^{\alpha\kappa}\epsilon^{\alg{a}\alg{b}}
\fL^\beta{}_\kappa
+\gamma\epsilon^{ab}\epsilon^{\alpha\beta}\epsilon^{\alg{a}\alg{k}}
\fC^\alg{b}{}_\alg{k}.
\end{align}
where the constants $\alpha$, $\beta$, $\gamma$ have to satisfy
$\alpha+\beta+\gamma=0$ due to the Jacobi identity.
Since the overall rescaling does not change the algebraic structure,
the only one relevant parameter which characterizes
$\alg{d}(2,1;\varepsilon)$ is $\varepsilon=-\gamma/\alpha$.
This exceptional algebra has a well-defined two-site quadratic Casimir
operator \eqref{casimir}, which enables many studies of the algebraic
properties \cite{MM}.
To reproduce the centrally extended superalgebra
$\alg{psu}(2|2)\ltimes{\mathbb R}^3$ without encountering the singular 
behavior, we shall rewrite
\begin{align}
\alpha=-1,\quad\beta=1-\varepsilon,\quad\gamma=\varepsilon,\quad
\fC^\alg{a}{}_\alg{b}
=\overline{\fC}{}^\alg{a}{}_\alg{b}/\varepsilon.
\end{align}
for the parameters $\alpha$, $\beta$, $\gamma$ and the last bosonic
$\alg{su}(2)$ generator $\fC^\alg{a}{}_\alg{b}$ (which becomes the
centers of $\alg{psu}(2|2)\ltimes{\mathbb R}^3$) and take the limit
$\varepsilon\to 0$.

Now let us regard each of three orthogonal sets of $\alg{su}(2)$ as
$\alg{so}(3)$ and define three kinds of three-dimensional gamma
matrices $(\gamma^{\bs{A}})^K{}_L=(\gamma^A_{A'})^K{}_L$ (where the
uppercase latin character $\bs{A}=(A,A')$ denotes either of the
lowercase latin character $\bs{a}=(a,a')$, the greek character
$\bs{\alpha}=(\alpha,\alpha')$ or the german character
$\bs{\alg{a}}=(\alg{a},\alg{a}')$) as
\begin{align}
(\gamma^{\bs{A}})^K{}_L=(\gamma^A_{A'})^K{}_L
=\frac{1}{\sqrt{2}}\epsilon^{KM}
(\epsilon_{MA'}\delta^A_L+\epsilon_{LA'}\delta^A_M)
=-\sqrt{2}\Bigl(\delta^K_{A'}\delta^A_L
-\frac{1}{2}\delta^K_L\delta^A_{A'}\Bigr).
\end{align}
Then, the gamma matrix satisfies the canonical Clifford algebra,
\begin{align}
(\gamma^{\bs{A}})^K{}_L(\gamma^{\bs{B}})^L{}_M
+(\gamma^{\bs{B}})^K{}_L(\gamma^{\bs{A}})^L{}_M
=2\delta^K_Mg^{\bs{A}\bs{B}},
\label{clifford}
\end{align}
with the metric defined as
\begin{align}
g^{\bs{A}\bs{B}}
=\delta^B_{A'}\delta^A_{B'}-\frac{1}{2}\delta^A_{A'}\delta^B_{B'},
\quad
g_{\bs{A}\bs{B}}
=\delta^{B'}_A\delta^{A'}_B-\frac{1}{2}\delta^{A'}_A\delta^{B'}_B,
\quad
\delta^{\bs{A}}_{\bs{B}}
=\delta^{B'}_{A'}\delta^A_B-\frac{1}{2}\delta^A_{A'}\delta^{B'}_B.
\end{align}
Note that in the metric we subtract the trace part, so that the
degree of freedom is $\delta^{\bs{A}}_{\bs{A}}=3$, which matches the
dimension of the adjoint representation of $\alg{su}(2)$.
Note also that
$(\gamma^{\bs{A}}\epsilon)^{KL}=(\gamma^{\bs{A}})^K{}_M\epsilon^{ML}$
and
$(\epsilon\gamma^{\bs{A}})_{KL}=\epsilon_{KM}(\gamma^{\bs{A}})^M{}_L$ 
are symmetric under the exchange of $K$ and $L$,
while the symmetry of the product of two gamma matrices reads
$(\gamma^{\bs{A}}\gamma^{\bs{B}}\epsilon)^{KL}
=-(\gamma^{\bs{B}}\gamma^{\bs{A}}\epsilon)^{LK}$.

Using these gamma matrices, the structure constants now take a simple
form
\begin{align}
f_{\fR^{\bs{c}}}{}^{\fR^{\bs{a}}\fR^{\bs{b}}}
&=\frac{1}{\sqrt{2}}
\Tr(\gamma_{\bs{c}}\gamma^{\bs{a}}\gamma^{\bs{b}}),
&
f_{\fF^{c\gamma\alg{c}}}{}^{\fR^{\bs{a}}\fF^{b\beta\alg{b}}}
&=\frac{-1}{\sqrt{2}}(\gamma^{\bs{a}})^b{}_c
\delta^\beta_\gamma\delta^\alg{b}_\alg{c},
&
f_{\fR^{\bs{c}}}{}^{\fF^{a\alpha\alg{a}}\fF^{b\beta\alg{b}}}
&=\frac{\alpha}{\sqrt{2}}(\gamma_{\bs{c}}\epsilon)^{ab}
\epsilon^{\alpha\beta}\epsilon^{\alg{a}\alg{b}},\nonumber\\
f_{\fL^{\bs{\gamma}}}{}^{\fL^{\bs{\alpha}}\fL^{\bs{\beta}}}
&=\frac{1}{\sqrt{2}}
\Tr(\gamma_{\bs{\gamma}}
\gamma^{\bs{\alpha}}\gamma^{\bs{\beta}}),
&
f_{\fF^{c\gamma\alg{c}}}{}^{\fL^{\bs{\alpha}}\fF^{b\beta\alg{b}}}
&=\frac{-1}{\sqrt{2}}(\gamma^{\bs{\alpha}})^\beta{}_\gamma
\delta^b_c\delta^\alg{b}_\alg{c},
&
f_{\fL^{\bs{\gamma}}}{}^{\fF^{a\alpha\alg{a}}\fF^{b\beta\alg{b}}}
&=\frac{\beta}{\sqrt{2}}(\gamma_{\bs{\gamma}}\epsilon)^{\alpha\beta}
\epsilon^{ab}\epsilon^{\alg{a}\alg{b}},\nonumber\\
f_{\fC^{\bs{\alg{c}}}}{}^{\fC^{\bs{\alg{a}}}\fC^{\bs{\alg{b}}}}
&=\frac{1}{\sqrt{2}}
\Tr(\gamma_{\bs{\alg{c}}}
\gamma^{\bs{\alg{a}}}\gamma^{\bs{\alg{b}}}),
&
f_{\fF^{c\gamma\alg{c}}}{}^{\fC^{\bs{\alg{a}}}\fF^{b\beta\alg{b}}}
&=\frac{-1}{\sqrt{2}}(\gamma^{\bs{\alg{a}}})^\alg{b}{}_\alg{c}
\delta^b_c\delta^\beta_\gamma,
&
f_{\fC^{\bs{\alg{c}}}}{}^{\fF^{a\alpha\alg{a}}\fF^{b\beta\alg{b}}}
&=\frac{\gamma}{\sqrt{2}}
(\gamma_{\bs{\alg{c}}}\epsilon)^{\alg{a}\alg{b}}
\epsilon^{ab}\epsilon^{\alpha\beta},
\end{align}
with those whose indices are lowered being
\begin{align}
f_{\fR^{\bs{a}}\fR^{\bs{b}}\fR^{\bs{c}}}
&=\frac{\alpha^2}{\sqrt{2}}
\Tr(\gamma_{\bs{a}}\gamma_{\bs{b}}\gamma_{\bs{c}}),
&
f_{\fF^{a\alpha\alg{a}}\fF^{b\beta\alg{b}}\fR^{\bs{c}}}
&=\frac{\alpha}{\sqrt{2}}(\epsilon\gamma_{\bs{c}})_{ab}
\epsilon_{\alpha\beta}\epsilon_{\alg{a}\alg{b}},\nonumber\\
f_{\fL^{\bs{\alpha}}\fL^{\bs{\beta}}\fL^{\bs{\gamma}}}
&=\frac{\beta^2}{\sqrt{2}}
\Tr(\gamma_{\bs{\alpha}}\gamma_{\bs{\beta}}\gamma_{\bs{\gamma}}),
&
f_{\fF^{a\alpha\alg{a}}\fF^{b\beta\alg{b}}\fL^{\bs{\gamma}}}
&=\frac{\beta}{\sqrt{2}}(\epsilon\gamma_{\bs{\gamma}})_{\alpha\beta}
\epsilon_{ab}\epsilon_{\alg{a}\alg{b}},\nonumber\\
f_{\fC^{\bs{\alg{a}}}\fC^{\bs{\alg{b}}}\fC^{\bs{\alg{c}}}}
&=\frac{\gamma^2}{\sqrt{2}}
\Tr(\gamma_{\bs{\alg{a}}}\gamma_{\bs{\alg{b}}}
\gamma_{\bs{\alg{c}}}),
&
f_{\fF^{a\alpha\alg{a}}\fF^{b\beta\alg{b}}\fC^{\bs{\alg{c}}}}
&=\frac{\gamma}{\sqrt{2}}
(\epsilon\gamma_{\bs{\alg{c}}})_{\alg{a}\alg{b}}
\epsilon_{ab}\epsilon_{\alpha\beta}.
\end{align}

\subsection{Representation}
An (infinite-dimensional) representation of the exceptional algebra
$\alg{d}(2,1;\varepsilon)$, which is a natural generalization of the
fundamental representation $\bs{2}|\bs{2}$ of
$\alg{psu}(2|2)\ltimes{\mathbb R}^3$, was constructed in \cite{MM}.
(See also \cite{VdJ,Bundyn}.)
Using the gamma matrices we have introduced, the representation is
given simply as,
\begin{align}
\fR^{\bs{a}}|\phi^k_m\rangle
&=-\frac{1}{\sqrt{2}}(\gamma^{\bs{a}})^k{}_l|\phi^l_m\rangle,
&
\fL^{\bs{\alpha}}|\psi^\kappa_r\rangle
&=-\frac{1}{\sqrt{2}}(\gamma^{\bs{\alpha}})^\kappa{}_\lambda
|\psi^\lambda_r\rangle,\nonumber\\
\fC^{\bs{\alg{a}}}|\phi^k_m\rangle
&=-\frac{1}{\sqrt{2}\gamma}
a^\alg{k}_m(\epsilon\gamma^{\bs{\alg{a}}})_{\alg{k}\alg{l}}
b^\alg{l}_{m+\bar{\alg{k}}+\bar{\alg{l}}}
|\phi^k_{m+\bar{\alg{k}}+\bar{\alg{l}}}\rangle,
&
\fC^{\bs{\alg{a}}}|\psi^\kappa_r\rangle
&=-\frac{1}{\sqrt{2}\gamma}
a^\alg{k}_{r+\bar{\alg{l}}}
(\epsilon\gamma^{\bs{\alg{a}}})_{\alg{k}\alg{l}}
b^\alg{l}_{r+\bar{\alg{l}}}
|\psi^\kappa_{r+\bar{\alg{k}}+\bar{\alg{l}}}\rangle,\nonumber\\
\fF^{a\alpha\alg{a}}|\phi^k_m\rangle
&=-a^\alg{a}_m\epsilon^{ka}|\psi^\alpha_{m+\bar{\alg{a}}}\rangle,
&
\fF^{a\alpha\alg{a}}|\psi^\kappa_r\rangle
&=b^\alg{a}_{r+\bar{\alg{a}}}
\epsilon^{\kappa\alpha}|\phi^a_{r+\bar{\alg{a}}}\rangle,
\label{exceptrep}
\end{align}
where the index of the bosonic state $m$ is an integer
$m\in{\mathbb Z}$,
while that of the fermionic state $r$ is a half-integer 
$r\in{\mathbb Z}+1/2$.
Also, $\bar{\alg{a}}=1/2$ (or $-1/2$) for $\alg{a}=1$ (or $2$
respectively).
Note that $\bar{\alg{a}}$ in the subscripts do not contract with any
of the superscripts.
Finally, $a^\alg{a}_m$ and $b^\alg{a}_m$ is defined as
\begin{align}
(a_m)^\alg{a}=\begin{pmatrix}a_m&-c_m\end{pmatrix},\quad
(b_m)^\alg{a}=\begin{pmatrix}b_m&-d_m\end{pmatrix},
\end{align}
subject to constraints
$\epsilon_{\alg{a}\alg{b}}a^\alg{a}_mb^\alg{b}_m=\alpha$
and $\epsilon_{\alg{a}\alg{b}}
a^\alg{a}_{r+\bar{\alg{b}}}b^\alg{b}_{r+\bar{\alg{b}}}=-\beta$
from the consistency of the algebra.

In the $\varepsilon\to 0$ limit, we find that the representation of
the algebra $\alg{psu}(2|2)\ltimes{\mathbb R}^3$ is given as
\begin{align}
\fR^{\bs{a}}|\phi^k\rangle
&=-\frac{1}{\sqrt{2}}(\gamma^{\bs{a}})^k{}_l|\phi^l\rangle,
&
\fL^{\bs{\alpha}}|\psi^\kappa\rangle
&=-\frac{1}{\sqrt{2}}(\gamma^{\bs{\alpha}})^\kappa{}_\lambda
|\psi^\lambda\rangle,
&
\overline{\fC}{}^{\bs{\alg{a}}}|\chi\rangle
&=\overline{C}{}^{\bs{\alg{a}}}|\chi\rangle,\nonumber\\
\fF^{a\alpha\alg{a}}|\phi^k\rangle
&=-a^\alg{a}\epsilon^{ka}|\psi^\alpha\rangle,
&
\fF^{a\alpha\alg{a}}|\psi^\kappa\rangle
&=b^\alg{a}\epsilon^{\kappa\alpha}|\phi^a\rangle,
\label{psurep}
\end{align}
where $|\chi\rangle$ denotes both bosonic and fermionic states and
$\overline{C}{}^{\bs{\alg{a}}}$ is defined as
\begin{align}
\overline{C}{}^{\bs{\alg{a}}}=-\frac{1}{\sqrt{2}}
a^\alg{k}(\epsilon\gamma^{\bs{\alg{a}}})_{\alg{k}\alg{l}}
b^\alg{l},
\quad
(\overline{\sC}\epsilon)^{\alg{b}\alg{c}}
=\frac{1}{\sqrt{2}}(a^\alg{b}b^\alg{c}+a^\alg{c}b^\alg{b}),
\end{align}
satisfying
$\overline{C}{}^{\bs{\alg{a}}}\overline{C}{}_{\bs{\alg{a}}}=1/2$.

\section{Serre relation}
Now let us start the proof that the evaluation representation is
compatible with the Serre relation.
As briefly mentioned in the introduction and explained more carefully
in appendix A, the origin of the Serre relation \eqref{serre} stems
from the homomorphism of the coproduct
\begin{align}
\Delta\bigl([\widehat\fJ^A,[\widehat\fJ^B,\fJ^C]]+\mbox{cyclic}\bigr)
=[\Delta\widehat\fJ^A,[\Delta\widehat\fJ^B,\Delta\fJ^C]]
+\mbox{cyclic},
\label{homomorphism}
\end{align}
and therefore plays an important role in discussing higher grade
generators.

After we reformulate the algebra and the representation in terms of
the gamma matrix in the previous section, using various formulas of
the gamma matrix, the computation of the right-hand-side of the Serre
relation \eqref{serre} now simply reduces to that of scattering
amplitudes of fermions.
Before we embark on the computation, let us make several remarks, some
of which will simplify the computation conceptually or technically.

First, let us note that the structure constant $f_{ABC}$ can be
regarded as the interaction among three particles $A$, $B$ and $C$
\cite{AS}.
Then, for example, the Jacobi identity \eqref{jacobi} is interpreted
as a relation stating that the summation of the scattering amplitudes
in the $s$-channel, $t$-channel and $u$-channel vanishes.
For a recent argument, see \cite{ABM}.

Secondly, to study the Serre relation \eqref{serre}, we have to
consider the product of four structure constants
$f_L{}^{AI}f_M{}^{BJ}f_N{}^{CK}f_{IJK}$.
Using the above interpretation, the contractions can be viewed as
particle interactions, where the initial states $A, B, C$ emit/absorb
the intermediate states $I, J, K$ (which interact among themselves)
and transit to the final states $L, M, N$.
Therefore, the contraction can be visualized in the Feynman diagram.
(See figure 1.)
We will separate our study into four cases depending on the number of
the fermionic generators $\fF$ in the initial states.

\begin{figure}[htb]
\begin{center}
\scalebox{1.0}[1.0]{\includegraphics{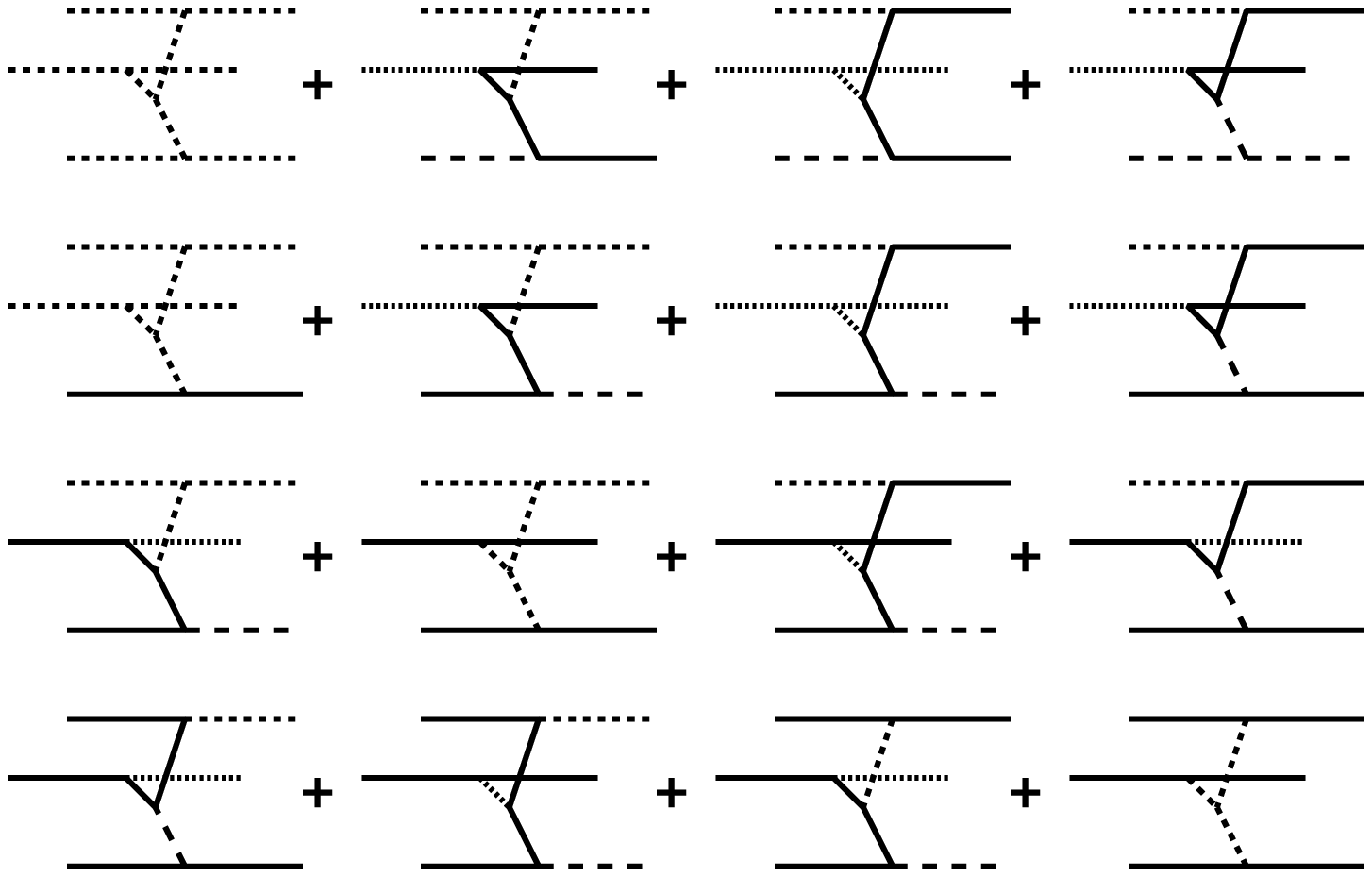}}
\setlength{\unitlength}{1mm}
\put(-147,88){\makebox(10,10){$\fB$}}
\put(-153,82){\makebox(10,10){$\fB$}}
\put(-147,72){\makebox(10,10){$\fB$}}
\put(-147,63){\makebox(10,10){$\fB$}}
\put(-153,56){\makebox(10,10){$\fB$}}
\put(-146,47){\makebox(10,10){$\fF$}}
\put(-147,37){\makebox(10,10){$\fB$}}
\put(-153,31){\makebox(10,10){$\fF$}}
\put(-146,21){\makebox(10,10){$\fF$}}
\put(-127,28){\makebox(10,10){$\fR/\fL/\overline\fC$}}
\put(-125,18){\makebox(10,10){$\fR/\fL/\overline\fC$}}
\put(-146,12){\makebox(10,10){$\fF$}}
\put(-153,6){\makebox(10,10){$\fF$}}
\put(-146,-4){\makebox(10,10){$\fF$}}
\put(-124,9){\makebox(10,10){$\fR/\fL/\overline\fC$}}
\put(-127,3){\makebox(10,10){$\fR/\fL/\overline\fC$}}
\put(-129,-2){\makebox(10,10){$\fR/\fL$}}
\put(-109,88){\makebox(10,10){$\fB$}}
\put(-115,83){\makebox(10,10){$\fB'$}}
\put(-110,72){\makebox(10,10){$\fB''$}}
\put(-109,63){\makebox(10,10){$\fB$}}
\put(-115,57){\makebox(10,10){$\fB'$}}
\put(-108,47){\makebox(10,10){$\fF$}}
\put(-87,44){\makebox(10,10){$\fR/\fL/\overline\fC$}}
\put(-109,37){\makebox(10,10){$\fB$}}
\put(-115,31){\makebox(10,10){$\fF$}}
\put(-108,21){\makebox(10,10){$\fF$}}
\put(-71,37){\makebox(10,10){$\fB$}}
\put(-76,31){\makebox(10,10){$\fF$}}
\put(-70,21){\makebox(10,10){$\fF$}}
\put(-65,28){\makebox(10,10){$\fR/\fL$}}
\put(-48,18){\makebox(10,10){$\fR/\fL/\overline\fC$}}
\put(-32,12){\makebox(10,10){$\fF$}}
\put(-38,6){\makebox(10,10){$\fF$}}
\put(-32,-4){\makebox(10,10){$\fF$}}
\put(-27,2){\makebox(10,10){$\fR/\fL$}}
\end{center}
\caption{The product of four structure constants
$f_L{}^{AI}f_M{}^{BJ}f_N{}^{CK}f_{IJK}$ in \eqref{serre} is
interpreted as particle interactions.
The initial states (left) emit/absorb the intermediate states (middle)
and transit to the final states (right).
Here the solid lines denote the fermionic generators while the various
dashed lines denote the various bosonic generators.
The bosonic generators in the initial states
$\fB,\fB',\fB''\in\{\fR,\fL\}$ are specified from the left-hand-side
of \eqref{serre}, though those in the intermediate or final states are
summed over $\fR$, $\fL$ and $\overline\fC$.
Each summation denotes the case with $0,1,2,3$ fermionic generators
$\fF$ in the initial states, respectively.}
\end{figure}

Thirdly, let us study the structure constants of
$\alg{d}(2,1;\varepsilon)$ more carefully.
There are only two types of structure constants:
$f_{\fF\fF\fB}\propto\epsilon\gamma\cdot\epsilon\cdot\epsilon$ and
$f_{\fB\fB\fB}\propto\Tr(\gamma\,\gamma\,\gamma)$ with $\fB$ standing for
any of the bosonic operators $\fR$, $\fL$ and $\fC$.
Every structure constant is factorized into three sectors corresponding
to three orthogonal $\alg{su}(2)$ subalgebras of
$\alg{d}(2,1;\varepsilon)$.
If we further denote indices of the adjoint representation of
$\alg{su}(2)$ by double lines and indices of the fundamental
representation by single lines, we can depict three auxiliary diagrams
for each original diagram as in figure 2.
These auxiliary diagrams enable us to study each sector separately.

\begin{figure}[htb]
\begin{center}
\scalebox{0.4}[0.4]{\includegraphics{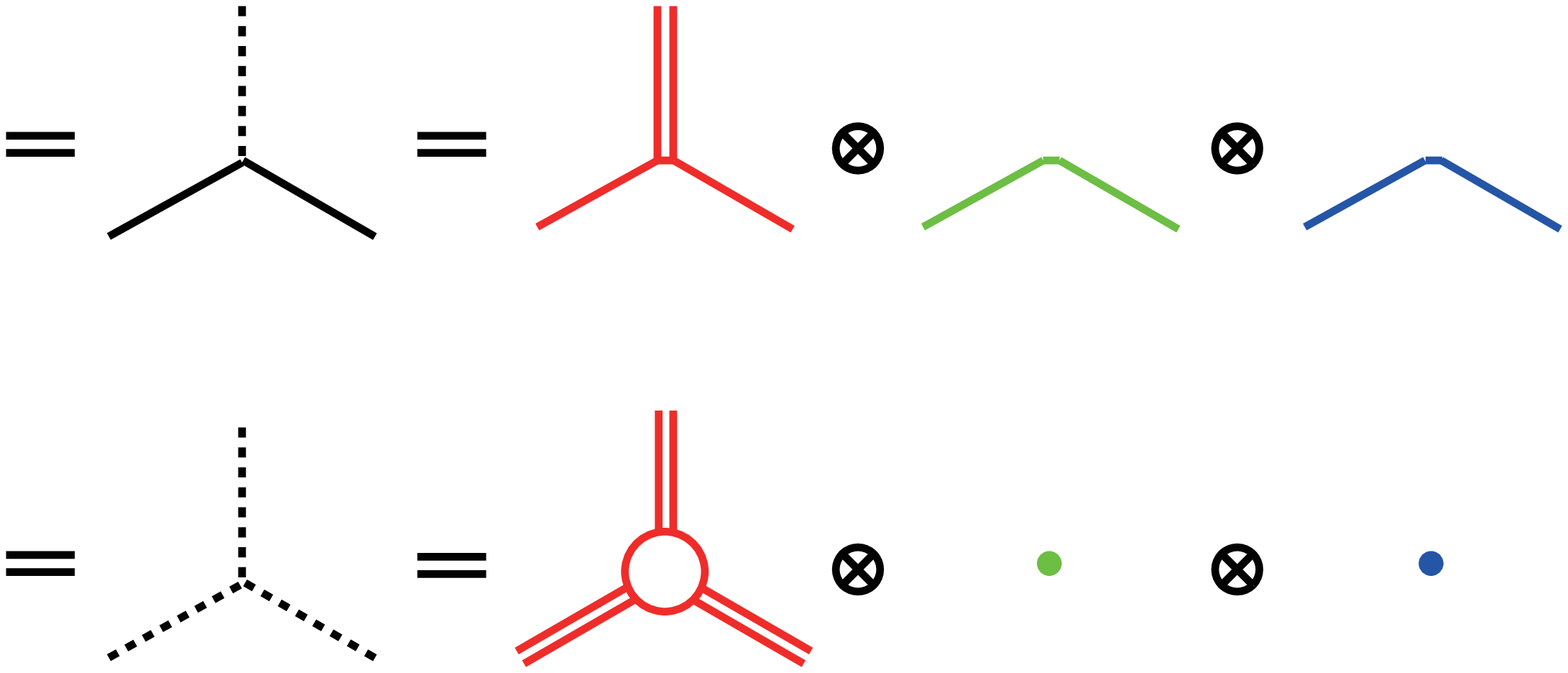}}
\setlength{\unitlength}{1mm}
\put(-100,23){\makebox(10,10){$f_{\fF\fF\fB}$}}
\put(-87,16){\makebox(10,10){$\fF$}}
\put(-72,16){\makebox(10,10){$\fF$}}
\put(-79,35){\makebox(10,10){$\fB$}}
\put(-100,0){\makebox(10,10){$f_{\fB\fB\fB}$}}
\put(-87,-7){\makebox(10,10){$\fB$}}
\put(-71,-7){\makebox(10,10){$\fB$}}
\put(-79,11){\makebox(10,10){$\fB$}}
\end{center}
\caption{Auxiliary diagrams are depicted for two types of the
structure constants, 
$f_{\fF\fF\fB}\propto\epsilon\gamma\cdot\epsilon\cdot\epsilon$ and 
$f_{\fB\fB\fB}\propto\Tr(\gamma\,\gamma\,\gamma)$ with
$\fB\in\{\fR,\fL,\fC\}$.
Since every structure constant is factorized into three sectors
corresponding to three orthogonal $\alg{su}(2)$ subalgebras of
$\alg{d}(2,1;\varepsilon)$, we can depict three auxiliary diagrams
with three different colors for the original diagram.
Here double lines denote indices of the adjoint representation while
single lines denote those of the fundamental representation.}
\end{figure}

Fourthly, we find that the contribution simply vanishes, if
$\overline{\fC}=\varepsilon\fC$ appears in the initial or intermediate
states.
This can be easily shown by noting the $\varepsilon$-dependence of the
structure constants,
\begin{align}
&f_{\overline{\fC}}{}^{\overline{\fC}\overline{\fC}}
=\cO(\varepsilon^1),\quad
f_\fF{}^{\overline{\fC}\fF}=\cO(\varepsilon^1),\quad
f_{\overline{\fC}}{}^{\fF\fF}=\cO(\varepsilon^0),\nonumber\\
&f_{\overline{\fC}\overline{\fC}\overline{\fC}}
=\cO(\varepsilon^{-1}),\quad
f_{\overline{\fC}\fF\fF}=\cO(\varepsilon^0).
\end{align}
If $\overline{\fC}$ appears in the initial or intermediate states,
$\overline{\fC}$ appears in the superscripts and we have to pick up a
factor of $\varepsilon^1$ which vanishes in the limit
$\varepsilon\to 0$, unless we use the singular structure constant
$f_{\overline{\fC}\overline{\fC}\overline{\fC}}
=\cO(\varepsilon^{-1})$.
If we use the singular structure constant, we need three structure
constants with $\overline{\fC}$ in the superscripts to contract all
the three subscripts, which result in the $\varepsilon$-dependence of
$\varepsilon^{-1}\varepsilon^3=\varepsilon^2$.

Finally, there is a ${\mathbb Z}_2$ symmetry.
If we exchange simultaneously the generators $\fR$ and $\fL$, the
parameters $a^\alg{a}$ and $b^\alg{a}$ and the states $|\phi\rangle$
and $|\psi\rangle$ (with an appropriate change of signs coming from
the parameters $\alpha$ and $\beta$), all the equations remains
correct.
Namely, it is enough to study half of the relations.

After all these remarks, let us start the computation.
In the calculation, it is useful to compute the totally symmetrized
generators $\{\fJ^L,\fJ^M,\fJ^N\}$ first, which we summarize in
appendix C.
Then, the computation can be easily performed with various formulas of
the gamma matrix collected in appendix B.

To summarize the results, we shall introduce here a new notation
\begin{align}
&\fJ^{(A_1A_2A_3)}_{\{C_1C_2C_3\}}
{\scriptstyle[B_1B_2B_3]}
=(-)^{B_1(A_2+B_2+A_3+B_3)+B_2(A_3+B_3)}\nonumber\\
&\qquad\qquad\qquad\times\{\fJ^{C_1},\fJ^{C_2},\fJ^{C_3}\}
f_{C_1}{}^{A_1B_1}f_{C_2}{}^{A_2B_2}f_{C_3}{}^{A_3B_3}f_{B_1B_2B_3}.
\end{align}
Hereafter we abbreviate the symbol $|A|$ introduced after
\eqref{jacobi} simply as $A$ when appearing in the exponents of the
sign $(-)$ to avoid unnecessary complications.

First, let us concentrate on the case with three bosonic generators in 
the initial states.
Since the initial states do not contain $\overline{\fC}$, the initial
states can be $(\fR,\fR,\fR)$, $(\fR,\fR,\fL)$, $(\fR,\fL,\fL)$,
$(\fL,\fL,\fL)$.
We start with the case that the initial states are $(\fR,\fR,\fR)$.
When acting on the bosonic state $|\phi\rangle$, the result vanishes
after summing over the cyclic combinations,
\begin{align}
\fJ^{(\fR\fR\fR)}_{\{\fR\fR\fR\}}
{\scriptstyle[\fR\fR\fR]}
|\phi\rangle=0,\quad
\bigl(\fJ^{(\fR\fR\fR)}_{\{\fR\fF\fF\}}
{\scriptstyle[\fR\fF\fF]}
+\fJ^{(\fR\fR\fR)}_{\{\fF\fR\fF\}}
{\scriptstyle[\fF\fR\fF]}
+\fJ^{(\fR\fR\fR)}_{\{\fF\fF\fR\}}
{\scriptstyle[\fF\fF\fR]}\bigr)
|\phi\rangle=0,
\end{align}
while when acting on the fermionic state $|\psi\rangle$, each term
vanishes independently.
Let us turn to the case with the initial states being
$(\fR,\fL,\fL)$.
Similarly, when acting on the bosonic state, the symmetric sum
vanishes,
\begin{align}
\fJ^{(\fR\fL\fL)}_{\{\fR\fF\fF\}}
{\scriptstyle[\fR\fF\fF]}|\phi\rangle=0,\quad
\bigl(\fJ^{(\fR\fL\fL)}_{\{\fF\fL\fF\}}
{\scriptstyle[\fF\fL\fF]}
+\fJ^{(\fR\fL\fL)}_{\{\fF\fF\fL\}}
{\scriptstyle[\fF\fF\fL]}\bigr)|\phi\rangle=0,
\end{align}
while when acting on the fermionic state, each term vanishes
separately.
The remaining initial states $(\fR,\fR,\fL)$ and $(\fL,\fL,\fL)$ can
also be seen from the above-mentioned ${\mathbb Z}_2$ symmetry.

Next, let us study the case with two bosonic generators in the initial
states.
Again due to the absence of $\overline{\fC}$ in the initial states, we 
have only to study three cases with the initial states being
$(\fR,\fR,\fF)$, $(\fR,\fL,\fF)$, $(\fL,\fL,\fF)$.
Let us start with the case with the initial states being
$(\fR,\fR,\fF)$.
This time, the following combinations vanish on any state
$|\chi\rangle$ including the bosonic one or the fermionic one.
\begin{align}
&\bigl(\fJ^{(\fR\fR\fF)}_{\{\fR\fR\fF\}}
{\scriptstyle[\fR\fR\fR]}
+\fJ^{(\fR\fR\fF)}_{\{\fR\fF\fR\}}
{\scriptstyle[\fR\fF\fF]}
+\fJ^{(\fR\fR\fF)}_{\{\fF\fR\fR\}}
{\scriptstyle[\fF\fR\fF]}\bigr)|\chi\rangle=0,\nonumber\\
&\bigl(\fJ^{(\fR\fR\fF)}_{\{\fR\fF\fL\}}
{\scriptstyle[\fR\fF\fF]}
+\fJ^{(\fR\fR\fF)}_{\{\fF\fR\fL\}}
{\scriptstyle[\fF\fR\fF]}\bigr)|\chi\rangle=0,\nonumber\\
&\bigl(\fJ^{(\fR\fR\fF)}_{\{\fR\fF\overline{\fC}\}}{\scriptstyle[\fR\fF\fF]}
+\fJ^{(\fR\fR\fF)}_{\{\fF\fR\overline{\fC}\}}{\scriptstyle[\fF\fR\fF]}
\bigr)|\chi\rangle=0,\nonumber\\
&\fJ^{(\fR\fR\fF)}_{\{\fF\fF\fF\}}
{\scriptstyle[\fF\fF\fR]}|\chi\rangle=0,\quad
\fJ^{(\fR\fR\fF)}_{\{\fF\fF\fF\}}
{\scriptstyle[\fF\fF\fL]}|\chi\rangle=0.
\end{align}
If the initial states are $(\fR,\fL,\fF)$, then all the contributions
are proportional to a single state with different coefficients.
Though it is difficult to specify which combinations of diagrams
cancel among themselves, we find it interesting to note that the pairs
with $\fR$ and $\fL$ almost exchanged cancel each other.
\begin{align}
&\bigl(\fJ^{(\fR\fL\fF)}_{\{\fR\fF\fR\}}{\scriptstyle[\fR\fF\fF]}
+\fJ^{(\fR\fL\fF)}_{\{\fF\fL\fL\}}{\scriptstyle[\fF\fL\fF]}\bigr)
|\chi\rangle=0,\nonumber\\
&\bigl(\fJ^{(\fR\fL\fF)}_{\{\fR\fF\fL\}}{\scriptstyle[\fR\fF\fF]}
+\fJ^{(\fR\fL\fF)}_{\{\fF\fL\fR\}}{\scriptstyle[\fF\fL\fF]}\bigr)
|\chi\rangle=0,\nonumber\\
&\bigl(
\fJ^{(\fR\fL\fF)}_{\{\fR\fF\overline{\fC}\}}{\scriptstyle[\fR\fF\fF]}
+\fJ^{(\fR\fL\fF)}_{\{\fF\fL\overline{\fC}\}}{\scriptstyle[\fF\fL\fF]}
\bigr)|\chi\rangle=0,\nonumber\\
&\fJ^{(\fR\fL\fF)}_{\{\fF\fF\fF\}}{\scriptstyle[\fF\fF\fR]}
|\chi\rangle=0,\quad
\fJ^{(\fR\fL\fF)}_{\{\fF\fF\fF\}}{\scriptstyle[\fF\fF\fL]}
|\chi\rangle=0.
\end{align}
Also, the case with $(\fL,\fL,\fF)$ is known from the ${\mathbb Z}_2$
symmetry.

Subsequently, the next subject is the case with only one bosonic
state, namely $(\fR,\fF,\fF)$ (with $(\fL,\fF,\fF)$ seen from the
above ${\mathbb Z}_2$ symmetry and $(\overline{\fC},\fF,\fF)$
vanishing by counting the $\varepsilon$-dependence).
When acting on the bosonic state, the result is proportional to a
single state again.
Since it is difficult to explain which combinations cancel among
themselves, we shall list up the results.
\begin{align}
&\Bigl(\fJ^{(\fR\fF\fF)}_{\{\fR\fR\fR\}}{\scriptstyle[\fR\fF\fF]}
+[\fJ^{(\fR\fF\fF)}_{\{\fR\fR\overline{\fC}\}}{\scriptstyle[\fR\fF\fF]}
+\mbox{sym}]
+\fJ^{(\fR\fF\fF)}_{\{\fR\overline{\fC}\overline{\fC}\}}
{\scriptstyle[\fR\fF\fF]}
+\fJ^{(\fR\fF\fF)}_{\{\fR\fF\fF\}}{\scriptstyle[\fR\fR\fR]}\nonumber\\
&+[\fJ^{(\fR\fF\fF)}_{\{\fF\fR\fF\}}{\scriptstyle[\fF\fF\fR]}+\mbox{sym}]
+[\fJ^{(\fR\fF\fF)}_{\{\fF\fL\fF\}}{\scriptstyle[\fF\fF\fR]}+\mbox{sym}]
+[\fJ^{(\fR\fF\fF)}_{\{\fF\overline{\fC}\fF\}}
{\scriptstyle[\fF\fF\fR]}+\mbox{sym}]\nonumber\\
&+[\fJ^{(\fR\fF\fF)}_{\{\fF\fR\fF\}}{\scriptstyle[\fF\fF\fL]}+\mbox{sym}]
+[\fJ^{(\fR\fF\fF)}_{\{\fF\fL\fF\}}{\scriptstyle[\fF\fF\fL]}+\mbox{sym}]
+[\fJ^{(\fR\fF\fF)}_{\{\fF\overline{\fC}\fF\}}
{\scriptstyle[\fF\fF\fL]}+\mbox{sym}]\Bigr)|\phi^k\rangle\nonumber\\
&=\Bigl(-\frac{5}{4}+0+\frac{3}{4}+1
+\frac{1}{4}-\frac{3}{8}-\frac{3}{8}
+\frac{3}{4}+\frac{3}{8}-\frac{9}{8}\Bigr)
\epsilon^{\beta\gamma}
\epsilon^{\alg{b}\alg{c}}
\Bigl[(\gamma^k_l\gamma^{\bs{a}}\epsilon)^{bc}
+(\gamma^k_l\gamma^{\bs{a}}\epsilon)^{cb}\Bigr]|\phi^l\rangle,
\end{align}
where sym denotes the symmetric combination under the exchange between
the second particle and the third one in the intermediate and final
states and we have specified the indices of the initial states as
$(\fR^{\bs{a}},\fF^{b\beta\alg{b}},\fF^{c\gamma\alg{c}})$.
When acting on the fermionic state, each symmetric combination in the
square parenthesis vanishes separately.

Finally, let us study the case without bosonic generators in the
initial state.
As in the case of $(\fR,\fL,\fF)$ there is a simple rule of the
cancellation.
We find that the diagrams with $\fR$ and $\fL$ exchanged cancel each
other.
Namely, the cancellation happens as follows.
\begin{align}
&\bigl(\fJ^{(\fF\fF\fF)}_{\{\fR\fR\fF\}}{\scriptstyle[\fF\fF\fR]}
+\fJ^{(\fF\fF\fF)}_{\{\fL\fL\fF\}}{\scriptstyle[\fF\fF\fL]}
+\mbox{cyc}\bigr)
|\chi\rangle=0,\nonumber\\
&\bigl(\fJ^{(\fF\fF\fF)}_{\{\fR\fR\fF\}}{\scriptstyle[\fF\fF\fL]}
+\fJ^{(\fF\fF\fF)}_{\{\fL\fL\fF\}}{\scriptstyle[\fF\fF\fR]}
+\mbox{cyc}\bigr)
|\chi\rangle=0,\nonumber\\
&\bigl(\fJ^{(\fF\fF\fF)}_{\{\fR\fL\fF\}}{\scriptstyle[\fF\fF\fR]}
+\fJ^{(\fF\fF\fF)}_{\{\fR\fL\fF\}}{\scriptstyle[\fF\fF\fL]}
+\mbox{tot sym}\bigr)
|\chi\rangle=0,\nonumber\\
&\bigl(\fJ^{(\fF\fF\fF)}_{\{\overline{\fC}\overline{\fC}\fF\}}
{\scriptstyle[\fF\fF\fR]}
+\fJ^{(\fF\fF\fF)}_{\{\overline{\fC}\overline{\fC}\fF\}}
{\scriptstyle[\fF\fF\fL]}
+\mbox{cyc}\bigr)
|\chi\rangle=0,\nonumber\\
&\bigl(\fJ^{(\fF\fF\fF)}_{\{\fR\overline{\fC}\fF\}}{\scriptstyle[\fF\fF\fR]}
+\fJ^{(\fF\fF\fF)}_{\{\fL\overline{\fC}\fF\}}{\scriptstyle[\fF\fF\fL]}
+\mbox{tot sym}\bigr)
|\chi\rangle=0,\nonumber\\
&\bigl(\fJ^{(\fF\fF\fF)}_{\{\fR\overline{\fC}\fF\}}{\scriptstyle[\fF\fF\fL]}
+\fJ^{(\fF\fF\fF)}_{\{\fL\overline{\fC}\fF\}}{\scriptstyle[\fF\fF\fR]}
+\mbox{tot sym}\bigr)
|\chi\rangle=0,\nonumber\\
&\bigl(\fJ^{(\fF\fF\fF)}_{\{\fF\fF\fF\}}{\scriptstyle[\fR\fR\fR]}
+\fJ^{(\fF\fF\fF)}_{\{\fF\fF\fF\}}{\scriptstyle[\fL\fL\fL]}\bigr)
|\chi\rangle=0,
\end{align}
where cyc denotes the summation over the cyclic rotation of three
particles in the intermediate and final states and tot sym denotes the
totally symmetric summation.

To summarize, we have proved that the right-hand-side of the Serre
relation \eqref{serre} vanishes on the representation, which enables
the fundamental representation of $\alg{psu}(2|2)\ltimes\mathbb{R}^3$
\eqref{psurep} to be the evaluation representation for the Yangian
algebra.

\section{Higher grade generators}
As we mentioned in the introduction, the meaning of the Serre relation
is to impose constraints on the higher grade generators \cite{MK}.
Namely, in terms of the BRST charge $\cQ$, the Serre relation
\eqref{serre}, stemming from the homomorphism of the coproduct
$\Delta\bigl(\cQ\bigl([\widehat\fJ,\widehat\fJ]\bigr)\bigr)
=\cQ\bigl([\Delta\widehat\fJ,\Delta\widehat\fJ]\bigr)$
\eqref{homomorphism}, takes the form
$\cQ\bigl([\widehat\fJ,\widehat\fJ]\bigr)=\fJ^3$ schematically, which
puts constraints on the commutator of the grade-1 generators
$[\widehat\fJ,\widehat\fJ]$.
As we have studied the Serre relation in the previous section, let us
now turn to the construction of higher grade generators.

It is natural to expect that the grade-2 generator is generated by the
commutator of the grade-1 generators
$[\widehat\fJ,\widehat\fJ]=\cQ\widehat{\widehat\fJ}+X$, namely,
\begin{align}
[\widehat\fJ^B,\widehat\fJ^C]
=\widehat{\widehat\fJ}{}^Af_A{}^{BC}+X^{BC}.
\label{grade2X}
\end{align}
The inclusion of an extra term $X^{BC}$ is inevitable from the
constraint of the Serre relation, which requires $X^{BC}$ to be
subject to the condition $\cQ X=\fJ^3$, or explicitly,
\begin{align}
X^{(A|D}f_D{}^{|BC)}
=\frac{\hbar^2}{24}(-)^{I(B+C)+JC}
\{\fJ^L,\fJ^M,\fJ^N\}f_L{}^{AI}f_M{}^{BJ}f_N{}^{CK}f_{KJI},
\label{serreX}
\end{align}
with the symbol $(A|\cdot|BC)$ in the superscripts denoting summation
over the cyclic rotation of $A$, $B$ and $C$ with the suitable
Grassmannian charges.
Note that there is a gauge ambiguity to shift $X$ and
$\widehat{\widehat{\fJ}}$ simultaneously by $X\to X-\cQ Y$ and
$\widehat{\widehat{\fJ}}\to\widehat{\widehat{\fJ}}+Y$. 

For the case that the quadratic Casimir operator of the adjoint
representation $c_2$ (defined by $f_A{}^{BC}f_{CBD}=c_2g_{AD}$) is
non-vanishing, there is a simple choice of the gauge fixing condition
$\cQ^{-1}X=0$, namely, $X^{BC}f_{CB}{}^A/c_2=0$.
In this gauge, we can solve \eqref{grade2X} for the grade-2 generator
and arrive at the canonical definition of it \cite{MK},
\begin{align}
\widehat{\widehat\fJ}{}^A
=\frac{1}{c_2}[\widehat\fJ^B,\widehat\fJ^C]f_{CB}{}^A,
\label{can_grade2}
\end{align}
Note that the expression \eqref{can_grade2} is a symmetry of the R-matrix
and the normalization is compatible with the evaluation representation
$\fJ^A_n\simeq(iu)^n\fJ^A$.
By plugging the coproduct of the grade-1 generator
\begin{align}
\Delta\widehat\fJ^A=\widehat\fJ^A\otimes 1+1\otimes\widehat\fJ^A
+\frac{\hbar}{2}\fJ^M\otimes\fJ^Nf_{NM}{}^A,
\label{coproduct1}
\end{align}
into the expression of the grade-2 generator \eqref{can_grade2}, the
coproduct of the grade-2 generator is found to be
\begin{align}
&\Delta\widehat{\widehat\fJ}{}^A
=\widehat{\widehat\fJ}{}^A\otimes 1
+1\otimes\widehat{\widehat\fJ}{}^A
+\frac{\hbar}{2}
\bigl(\widehat\fJ^M\otimes\fJ^N
+\fJ^M\otimes\widehat\fJ^N\bigr)
f_{NM}{}^A\nonumber\\
&\quad -\frac{\hbar^2}{8c_2}(-)^{IC}
\bigl(\{\fJ^L,\fJ^M\}\otimes\fJ^N
+(-)^{N(L+M)}\fJ^N\otimes\{\fJ^L,\fJ^M\}\bigr)
f_{L}{}^{BI}f_{M}{}^{CJ}f_{NJI}f_{CB}{}^A.
\label{can_coprd2}
\end{align}

Note that the results \eqref{can_grade2} and \eqref{can_coprd2} rely
heavily on the fact that the quadratic Casimir operator $c_2$ is
non-vanishing.
However, for the superalgebra we are considering now it is not
difficult to show $c_2=0$ even for the $\varepsilon$-deformed
exceptional superalgebra $\alg{d}(2,1;\varepsilon)$.
In this case, we have to take another suitable gauge fixing
condition.
To respect the compatibility of the evaluation representation in
\eqref{grade2X}, the extra term $X^{BC}$ must be vanishing on all
states.
Therefore, it is natural to choose a gauge so that
\begin{align}
X^{BC}|\chi\rangle=0.
\label{gauge}
\end{align}
Then the problem is reduced to find an expression of $X^{BC}$
which satisfies the Serre relation \eqref{serreX} and the above gauge
condition \eqref{gauge} simultaneously.
A naive candidate coming from the derivation of the Serre relation in
appendix A is (See e.g.\ \eqref{Hint}.)
\begin{align}
\widetilde X^{BC}
=-\frac{1}{3}\cdot\frac{\hbar^2}{24}(-)^{IC}
\{\fJ^L,\fJ^M,\fJ^N\}
f_L{}^{BI}f_M{}^{CJ}f_{NJI}.
\label{Xdef}
\end{align}
As the computation in appendix A, using the Jacobi identity twice, we
can show that this expression $\widetilde X^{BC}$ is a solution to
\eqref{serreX}.
However, this is not compatible with the condition \eqref{gauge}.
Actually, using the formulas summarized in appendices B and C, it is
not difficult to show that $\widetilde X^{BC}$ takes the following
form on the fundamental representation of
$\alg{psu}(2|2)\ltimes\mathbb{R}^3$,
\begin{align}
\widetilde X^{BC}|\chi\rangle
=-2\cdot\frac{\hbar^2}{24}[\fJ^B,\fJ^C]|\chi\rangle.
\label{Xpsu}
\end{align}
In order to satisfy the condition \eqref{gauge}, we can improve
$\widetilde{X}$ without violating the Serre relation \eqref{serreX} by
subtracting the commutator since it is a $\cQ$-exact term, or
schematically $\widetilde X\to\widetilde X-\cQ Y$.
Finally we adopt the following expression for $X^{BC}$,
\begin{align}
X^{BC}
=-\frac{1}{3}\cdot\frac{\hbar^2}{24}
\Bigl((-)^{IC}\{\fJ^L,\fJ^M,\fJ^N\}
f_L{}^{BI}f_M{}^{CJ}f_{NJI}
-6\fJ^Af_A{}^{BC}\Bigr).
\label{X}
\end{align}
Note that the relation \eqref{Xpsu} is interesting by itself.
As mentioned above, $\widetilde{X}^{(A|D}f_D{}^{|BC)}$ (or
$\cQ\widetilde{X}$) is equal to the right-hand-side of the Serre
relation \eqref{serre} at the algebraic level.
Since \eqref{Xpsu} indicates that $\widetilde{X}$ is $\cQ$-exact on
the representation, this gives an alternative proof of the
compatibility of the evaluation representation with the Serre
relation, which has been directly proved in section 3.
This argument will be important in the next section.

As the final project of this section, let us construct the coproduct
of the grade-2 generator by requiring the homomorphism of the
coproduct for the grade-1 commutator
$\Delta\bigl([\widehat\fJ^A,\widehat\fJ^B]\bigr)
=[\Delta\widehat\fJ{}^A,\Delta\widehat\fJ{}^B]$.
Using the commutator of the grade-1 generators as in \eqref{grade2X},
we can write down the coproduct coupled with the structure constant
as,
\begin{align}
\Delta\widehat{\widehat\fJ}{}^Af_A{}^{BC}
=[\Delta\widehat\fJ^B,\Delta\widehat\fJ^C]
-\Delta X^{BC}.
\label{def_grade2coprd}
\end{align}

The reader may worry that the quadratic Casimir operator $c_2$ may
appear again on the left-hand-side when we solve for the grade-2
generator $\Delta\widehat{\widehat\fJ}{}^A$.
However, this is not the case.
Because it can be shown that the right-hand-side is also proportional
to the structure constants $f_A{}^{BC}$, we find a solution of the
grade-2 generator just by dropping the same structure constants.
More precisely, the vanishing of $c_2$ in the superalgebra
$\alg{d}(2,1;\varepsilon)$ 
(and its limiting case $\alg{psu}(2|2)\ltimes{\mathbb R}^3)$ is a
consequence of summing over all the bosonic and fermionic intermediate
generators $B$ and $C$ in $f_A{}^{BC}f_{CBD}=c_2g_{AD}$.
However, if we fix the intermediate generators and sum only over all
the indices of the generators, the result is proportional to the
Killing form $g_{AD}$ with a non-vanishing coefficient,
\begin{align}
f_{\fR^{\bs{a}}}{}^{\fR^{\bs{b}}\fR^{\bs{c}}}
f_{\fR^{\bs{c}}\fR^{\bs{b}}\fR^{\bs{d}}}
&=-4\alpha g_{\fR^{\bs{a}}\fR^{\bs{d}}},&
f_{\fR^{\bs{a}}}{}^{\fF^B\fF^C}
f_{\fF^C\fF^B\fR^{\bs{d}}}
&=4\alpha g_{\fR^{\bs{a}}\fR^{\bs{d}}},&
f_{\fF^A}{}^{\fR^{\bs{b}}\fF^C}
f_{\fF^C\fR^{\bs{b}}\fF^D}
&=-\frac{3}{2}\alpha g_{\fF^A\fF^D},\nonumber\\
f_{\fL^{\bs{\alpha}}}{}^{\fL^{\bs{\beta}}\fL^{\bs{\gamma}}}
f_{\fL^{\bs{\gamma}}\fL^{\bs{\beta}}\fL^{\bs{\delta}}}
&=-4\beta g_{\fL^{\bs{\alpha}}\fL^{\bs{\delta}}},&
f_{\fL^{\bs{\alpha}}}{}^{\fF^B\fF^C}
f_{\fF^C\fF^B\fL^{\bs{\delta}}}
&=4\beta g_{\fL^{\bs{\alpha}}\fL^{\bs{\delta}}},&
f_{\fF^A}{}^{\fL^{\bs{\beta}}\fF^C}
f_{\fF^C\fL^{\bs{\beta}}\fF^D}
&=-\frac{3}{2}\beta g_{\fF^A\fF^D},\nonumber\\
f_{\fC^{\bs{\alg{a}}}}{}^{\fC^{\bs{\alg{b}}}\fC^{\bs{\alg{c}}}}
f_{\fC^{\bs{\alg{c}}}\fC^{\bs{\alg{b}}}\fC^{\bs{\alg{d}}}}
&=-4\gamma g_{\fC^{\bs{\alg{a}}}\fC^{\bs{\alg{d}}}},&
f_{\fC^{\bs{\alg{a}}}}{}^{\fF^B\fF^C}
f_{\fF^C\fF^B\fC^{\bs{\alg{d}}}}
&=4\gamma g_{\fC^{\bs{\alg{a}}}\fC^{\bs{\alg{d}}}},&
f_{\fF^A}{}^{\fC^{\bs{\alg{b}}}\fF^C}
f_{\fF^C\fC^{\bs{\alg{b}}}\fF^D}
&=-\frac{3}{2}\gamma g_{\fF^A\fF^D},
\end{align}
where we have introduced the capital Latin letter $A$ temporally to
represent all of the fermionic indices $(a,\alpha,\alg{a})$ for the
notational simplicity.
The Killing form of the bosonic generators is proportional to the
metric
$g_{\fR^{\bs{a}}\fR^{\bs{b}}}
=-\alpha g_{\bs{a}\bs{b}}$,
$g_{\fL^{\bs{\alpha}}\fL^{\bs{\beta}}}
=-\beta g_{\bs{\alpha}\bs{\beta}}$,
$g_{\fC^{\bs{\alg{a}}}\fC^{\bs{\alg{b}}}}
=-\gamma g_{\bs{\alg{a}}\bs{\alg{b}}}$,
while that of the fermionic ones are
$g_{\fF^A\fF^B}
=\epsilon_{ab}\epsilon_{\alpha\beta}\epsilon_{\alg{a}\alg{b}}$.
Therefore, when the right-hand-side of \eqref{def_grade2coprd} is
proportional to the structure constant $f_A{}^{BC}$, we can restrict
the intermediate generators $B$ and $C$ to specific generators and
obtain the grade-2 generator without using the quadratic Casimir
operator $c_2$.

After all, the explicit form of the coproduct is
\begin{align}
&\Delta\widehat{\widehat\fJ}{}^A
=\widehat{\widehat\fJ}{}^A\otimes 1
+1\otimes\widehat{\widehat\fJ}{}^A
+\frac{\hbar}{2}
\bigl(\widehat\fJ^M\otimes\fJ^N
+\fJ^M\otimes\widehat\fJ^N\bigr)
f_{NM}{}^A\nonumber\\
&
\qquad\quad -\frac{\hbar^2}{24}
\bigl(\{\fJ^L,\fJ^M\}\otimes\fJ^N
+(-)^{N(L+M)}\fJ^N\otimes\{\fJ^L,\fJ^M\}\bigr)
f_L{}^{AI}f_{MNI},
\label{coprd2}
\end{align}
which is obviously the symmetry of the R-matrix by the construction
\eqref{def_grade2coprd} and valid even if $c_2=0$.

\section{Lift to exceptional superalgebra}
In our discussion of the higher grade generators in the previous
section, we have found a simple expression \eqref{serreX} for the
right-hand-side of the Serre relation in terms of $\widetilde X^{BC}$
\eqref{Xdef}.
Namely, the compatibility of the evaluation representation reduces to
the equation
\begin{align}
\widetilde X^{(A|D}f_D{}^{|BC)}|\chi\rangle=0,
\label{Xf}
\end{align}
when acting on the representation.
For the previous case of $\alg{psu}(2|2)\ltimes{\mathbb R}^3$, it is
surprising to find that actually
\eqref{Xpsu} holds.
Then, the compatibility \eqref{Xf} is automatic from the nilpotency of
the BRST charge $\cQ$ (or the Jacobi identity).
In this simplified expression, it is not difficult to find that the
above argument is still valid for the case of the exceptional
superalgebra $\alg{d}(2,1;\varepsilon)$.
Namely, using the formulas in appendix D for
$\alg{d}(2,1;\varepsilon)$, after some computation we find a relation
which is a natural lift of \eqref{Xpsu},
\begin{align}
\widetilde X^{BC}|\chi\rangle
=2\alpha\beta\cdot\frac{\hbar^2}{24}[\fJ^B,\fJ^C]|\chi\rangle,
\label{Xd}
\end{align}
where the indices $B$ and $C$ run over all generators including
$\fC^{\bs{\alg{a}}}$.
This final result implies that the infinite-dimensional representation
of $\alg{d}(2,1;\varepsilon)$ \eqref{exceptrep} is also extended to
the evaluation representation of the Yangian algebra.
Since the expression is parallel to the case of
$\alg{psu}(2|2)\ltimes{\mathbb R}^3$, the higher grade generators can
be constructed similarly as in the previous section.

\section{Discussion}
We have proved that the evaluation representation adopted in the
AdS/CFT spin chain model is compatible with the Serre relation of the
Yangian algebra for both the cases of
$\alg{psu}(2|2)\ltimes{\mathbb R}^3$ and $\alg{d}(2,1;\varepsilon)$.
Conceptually, the Serre relation imposes constraints on higher grade
generators.
Therefore, it is inevitable to prove the Serre relation in order to
proceed to the infinite symmetry of string theory.
Technically, we have introduced a new formalism for the algebra and
the representation via the three-dimensional gamma matrix.
We have also found that the generator $\widetilde X^{BC}$ coming from
later discussions on higher grade generators is useful to simplify the 
proof.
We believe that our formulation can further apply to many related
computations as well.
For example, the determination of the R-matrix for the fundamental
excitations \cite{BRmat,T} and the bound states \cite{bound} should
be done in a much simpler way using our gamma matrix formulation.

We also analyze the higher grade generators, where we find that the
argument and the formula become subtle and singular because the
quadratic Casimir operator of the adjoint representation $c_2$ is
vanishing.
Here by changing the gauge fixing condition we propose an alternative
non-singular construction of the higher grade generator which is a
symmetry of the R-matrix and at the same time compatible with the
evaluation representation.

Let us list several further directions to conclude this paper.

As we have suggested in the introduction, the compatibility of the
evaluation representation may indicate a close relation to the string
worldsheet theory.
The fact that the evaluation representation is lifted to the
exceptional superalgebra $\alg{d}(2,1;\varepsilon)$ may imply its
string interpretation.
We would like to pursue this possibility further in our future work.

In our simplified proof, we introduce a generator $\widetilde{X}^{BC}$
and find that it is proportional to the commutator $[\fJ^B,\fJ^C]$ on
the representation.
Since it can be seen directly from the definition \eqref{Xdef} that
$\widetilde{X}{}^{BC}$ is antisymmetric (in the generalized sense) and
the result has to transform canonically as the product of $\fJ^B$ and
$\fJ^C$, it is not surprising to find the proportionality.
What is surprising is that all of the results come with the same
proportionality constant.
Our understanding would be clearer if we can rewrite the
proportionality constant $-2$ and $2\alpha\beta$ in \eqref{Xpsu} and
\eqref{Xd} in the terminology of the representation theory.
The value of the quadratic Casimir operator for
$\alg{d}(2,1;\varepsilon)$,
$\cT^\alg{d}|\chi\rangle=(\alpha\beta/4\gamma)|\chi\rangle$
would be a clue.

A topic related to our motivation of constructing higher grade
generators is the universal R-matrix \cite{universal, ST}, where the
Chevalley-Serre basis is conventionally adopted.
It would be great to find the universal R-matrix both for
$\alg{psu}(2|2)\ltimes{\mathbb R}^3$ and $\alg{d}(2,1;\varepsilon)$,
which contains all of the higher grade generators.

An interesting proposal that the conventional and dual superconformal
symmetries form the Yangian algebra together was made very recently
\cite{DHP}, where the vanishing Casimir operator $c_2=0$ plays an
important role in the consistency.
We would like to understand the relation to our present discussion.

Since the interacting theory of string theory is not so different from 
the free theory locally on the worldsheet and since the correspondence
between gauge theory and string theory works well also at the
interacting level \cite{SFT,CJJK}, it would be interesting to study
the integrability at the interacting level more extensively and define
interacting string theory with Yang-Mills theory.

\section*{Acknowledgments}
We are grateful to H.~Awata, M.~Hatsuda, A.~Ishida, H.~Kanno,
J.~Gomis, T.~Nakanishi, K.~Sakai, K.~Takahashi, S.~Teraguchi,
A.~Torrielli, A.~Tsuchiya and T.~Yamashita for valuable discussions.
One of the authors (S.M.) is also grateful for Yukawa Institute for
Theoretical Physics at Kyoto University for hospitality, where part of
this work was done.
The work of S.M. is supported partly by Grant-in-Aid from Daiko
Foundation and partly by Grant-in-Aid for Young Scientists (B)
[\#18740143] and [\#21740176] from the Japan Ministry of Education,
Culture, Sports, Science and Technology.
The work of T.M. is supported partly by the Grant-in-Aid for the GCOE
program ``Quest for Fundamental Principles in the Universe''.

\appendix
\section{Homomorphism of coproduct}
In this appendix we would like to explain that the origin of the Serre
relation stems from the homomorphism of the coproduct of the grade-1 
generators defined in \eqref{coproduct1}.
The result in this appendix is well-known and also appears previously
in, for example, \cite{MK,DNW}.
The reason we recapitulate it here is because of the possible sign
ambiguity in dealing with the superalgebra.
In sections 3-5 we have found many cancellations and simplifications,
which do not happen if we take the wrong signs.

Let us start with computing 
$[\Delta\widehat\fJ^A,[\Delta\widehat\fJ^B,\Delta\fJ^C]]
=[\Delta\widehat\fJ^A,\Delta\widehat\fJ^Jf_J{}^{BC}]$.
{}From the expression
\begin{align}
[\fJ^P\otimes\fJ^Q,\fJ^R\otimes\fJ^S]
=\frac{1}{2}(-)^{QR}\bigl([\fJ^P,\fJ^R]\otimes\{\fJ^Q,\fJ^S\}
+\{\fJ^P,\fJ^R\}\otimes[\fJ^Q,\fJ^S]\bigr),
\end{align}
we find that the terms with only Lie algebra generators are given as
\begin{align}
[\Delta\widehat\fJ^A,[\Delta\widehat\fJ^B,\Delta\fJ^C]]
\Big|_{{\rm Lie}}
&=\frac{\hbar^2}{8}
(-)^{IL+IM+KM}f_{LI}{}^Af_{MK}{}^Jf_N{}^{IK}f_J{}^{BC}
\nonumber\\
&\quad\times\bigl(\{\fJ^L,\fJ^M\}\otimes\fJ^N
+(-)^{N(L+M)}\fJ^N\otimes\{\fJ^L,\fJ^M\}\bigr).
\label{Hint}
\end{align}
Here the signs can be interpreted as the Grassmannian charges when we
contract the superscripts with the subscripts in the canonical
position of the spinor indices as in \eqref{serre}.
We can continue the computations keeping the above interpretation of
the signs systematically.
For this purpose, we have to introduce an extra sign
$(-)^{(P+Q)(R+S+T)}$, even if we simply commute $\fJ^Nf_N{}^{PQ}$ with
the bosonic structure constants $f^{RST}$.
Further, using twice the Jacobi identity 
\begin{align}
f_{MK}{}^Jf_J{}^{BC}&=-(-)^{K(B+C)}f_M{}^{BJ}f_J{}^C{}_K
-(-)^{C(K+B)}f_M{}^{CJ}f_{JK}{}^B,\nonumber\\
f_{NI}{}^Kf_{KJ}{}^C&=-(-)^{C(I+J)}f_N{}^{CK}f_{KIJ}
-(-)^{I(J+C)}f_{NJ}{}^Kf_K{}^C{}_I,
\end{align}
and noting
$\{\fJ^L,\fJ^M\}f_L{}^{AI}f_M{}^{CJ}
=(-)^{(A+I)(C+J)}\{\fJ^L,\fJ^M\}f_L{}^{CJ}f_M{}^{AI}$,
finally we arrive at the expression
\begin{align}
&[\Delta\widehat\fJ^A,[\Delta\widehat\fJ^B,\Delta\fJ^C]]
\Big|_{{\rm Lie}}
=\frac{\hbar^2}{8}\bigl(\{\fJ^L,\fJ^M\}\otimes\fJ^N
+(-)^{N(L+M)}\fJ^N\otimes\{\fJ^L,\fJ^M\}\bigr)\nonumber\\
&\quad\times
\bigl((-)^{JC+I(B+C)}f_L{}^{AI}f_M{}^{BJ}f_N{}^{CK}f_{KJI}
\nonumber\\
&\qquad+(-)^{IB}f_L{}^{AI}f_M{}^{BJ}f_{NJ}{}^Kf_{KI}{}^C
-(-)^{IA+C(A+B)}f_L{}^{CI}f_M{}^{AJ}f_{NJ}{}^Kf_{KI}{}^B\bigr).
\end{align}
Adding the terms coming from the cyclic rotation of $(A,B,C)$, the
second term and the third term cancel each other while the first term
becomes totally symmetric in the indices $L, M, N$.
The final form indicates that the mapping of coproduct $\Delta$ is
homomorphic with respect to the bracket product \eqref{homomorphism}
if the Serre relation holds.

\section{Useful formulas of gamma matrices}
Here we collect some of the formulas of the gamma matrices.
First using the definition of the gamma matrices \eqref{clifford}, we
can show the following ones without difficulty.
\begin{align}
&(\gamma^{\bs{A}}\gamma_{\bs{A}})^K{}_L=3\delta^K_L,\quad
(\gamma^{\bs{B}}\gamma^{\bs{A}}\gamma_{\bs{B}})^K{}_L
=-(\gamma^{\bs{A}})^K{}_L,\quad
(\gamma^{\bs{C}}\gamma^{\bs{A}}\gamma^{\bs{B}}\gamma_{\bs{C}})^K{}_L
=(\gamma^{\bs{A}}\gamma^{\bs{B}}
+2\gamma^{\bs{B}}\gamma^{\bs{A}})^K{}_L,\nonumber\\
&\Tr(\gamma^{\bs{A}}\gamma^{\bs{B}})=2g^{\bs{A}\bs{B}},\quad
\Tr(\gamma^{\bs{A}}\gamma^{\bs{B}}\gamma^{\bs{C}}\gamma^{\bs{D}})
=2(g^{\bs{A}\bs{B}}g^{\bs{C}\bs{D}}
-g^{\bs{A}\bs{C}}g^{\bs{B}\bs{D}}
+g^{\bs{A}\bs{D}}g^{\bs{B}\bs{C}}).
\end{align}
If we further combine the use of the Fierz identity
\begin{align}
(\gamma^{\bs{A}})^K{}_L(\gamma_{\bs{A}})^M{}_N
=2\delta^K_N\delta^M_L-\delta^K_L\delta^M_N
=-\sqrt{2}(\gamma^K_L)^M{}_N,
\end{align}
we also find that
\begin{align}
&\Tr(\gamma^{\bs{A}}\gamma^{\bs{B}}\gamma^{\bs{C}})
(\gamma_{\bs{C}})^K{}_L
=2((\gamma^{\bs{A}}\gamma^{\bs{B}})^K{}_L
-\delta^K_Lg^{\bs{A}\bs{B}}),\nonumber\\
&\Tr(\gamma^{\bs{A}}\gamma^{\bs{B}}\gamma^{\bs{E}})
\Tr(\gamma^{\bs{C}}\gamma^{\bs{D}}\gamma_{\bs{E}})
=4(-g^{\bs{A}\bs{C}}g^{\bs{B}\bs{D}}
+g^{\bs{A}\bs{D}}g^{\bs{B}\bs{C}}).
\end{align}
Note that the last formula is nothing but the famous one
$\epsilon_{\bs{A}\bs{B}\bs{E}}\epsilon_{\bs{C}\bs{D}\bs{E}}
=\delta_{\bs{A}\bs{C}}\delta_{\bs{B}\bs{D}}
-\delta_{\bs{A}\bs{D}}\delta_{\bs{B}\bs{C}}$ for
$\bs{A},\bs{B},\cdots=1,2,3$.

\section{Totally symmetrized generators}
It is useful here to summarize the action of various totally
symmetrized generators $\{\fJ^A,\fJ^B,\fJ^C\}$ appearing in the
computation.
We classify the formulas into four cases depending on whether the
totally symmetrized generators and the states are bosonic or
fermionic.

$\bullet\;\{$bosonic$\}|\phi\rangle$
\begin{align}
&
\{\fR^{\bs{a}},\fR^{\bs{b}},\fR^{\bs{c}}\}|\phi^k\rangle
=-\frac{1}{\sqrt{2}}(g^{\bs{b}\bs{c}}\gamma^{\bs{a}}
+g^{\bs{c}\bs{a}}\gamma^{\bs{b}}
+g^{\bs{a}\bs{b}}\gamma^{\bs{c}})^k{}_l
|\phi^l\rangle,\nonumber\\
&
\{\fR^{\bs{a}},\fR^{\bs{b}},
\overline{\fC}{}^{\bs{\alg{c}}}\}|\phi^k\rangle
=3\overline{C}{}^{\bs{\alg{c}}}g^{\bs{a}\bs{b}}|\phi^k\rangle,\quad
\{\fR^{\bs{a}},\overline{\fC}{}^{\bs{\alg{b}}},
\overline{\fC}{}^{\bs{\alg{c}}}\}|\phi^k\rangle
=-3\sqrt{2}\overline{C}^{\bs{\alg{b}}}\overline{C}^{\bs{\alg{c}}}
(\gamma^{\bs{a}})^k{}_l|\phi^l\rangle,\nonumber\\
&
\{\fR^{\bs{a}},\fF^{b\beta\alg{b}},\fF^{c\gamma\alg{c}}\}
|\phi^k\rangle
=-\epsilon^{\beta\gamma}
\Bigl[(\gamma^{\bs{a}}\epsilon)^{bc}\delta^k_l
(\overline{\sC}\epsilon)^{\alg{b}\alg{c}}
+\frac{1}{\sqrt{2}}\epsilon^{bc}(\gamma^{\bs{a}})^k{}_l
\epsilon^{\alg{b}\alg{c}}\Bigr]
|\phi^l\rangle,\nonumber\\
&
\{\fL^{\bs{\alpha}},\fF^{b\beta\alg{b}},\fF^{c\gamma\alg{c}}\}
|\phi^k\rangle
=\frac{1}{2}(\gamma^{\bs{\alpha}}\epsilon)^{\beta\gamma}
\Bigl[\epsilon^{bc}\delta^k_l
(\overline{\sC}\epsilon)^{\alg{b}\alg{c}}
-(\gamma^k_l\epsilon)^{bc}\epsilon^{\alg{b}\alg{c}}\Bigr]
|\phi^l\rangle,\nonumber\\
&
\{\overline{\fC}{}^{\bs{\alg{a}}},
\fF^{b\beta\alg{b}},\fF^{c\gamma\alg{c}}\}|\phi^k\rangle
=-\frac{3}{2}\overline{C}{}^{\bs{\alg{a}}}\epsilon^{\beta\gamma}
\Bigl[2(\gamma^k_l\epsilon)^{bc}
(\overline{\sC}\epsilon)^{\alg{b}\alg{c}}
-\epsilon^{bc}\delta^k_l\epsilon^{\alg{b}\alg{c}}\Bigr]
|\phi^l\rangle.
\label{Bphi}
\end{align}

$\bullet\;\{$bosonic$\}|\psi\rangle$
\begin{align}
&
\{\fL^{\bs{\alpha}},\fL^{\bs{\beta}},\fL^{\bs{\gamma}}\}
|\psi^\kappa\rangle
=-\frac{1}{\sqrt{2}}(g^{\bs{\beta}\bs{\gamma}}\gamma^{\bs{\alpha}}
+g^{\bs{\gamma}\bs{\alpha}}\gamma^{\bs{\beta}}
+g^{\bs{\alpha}\bs{\beta}}\gamma^{\bs{\gamma}})^\kappa{}_\lambda
|\psi^\lambda\rangle,\nonumber\\
&
\{\fL^{\bs{\alpha}},\fL^{\bs{\beta}},
\overline{\fC}{}^{\bs{\alg{c}}}\}|\psi^\kappa\rangle
=3\overline{C}{}^{\bs{\alg{c}}}g^{\bs{\alpha}\bs{\beta}}
|\psi^\kappa\rangle,\quad
\{\fL^{\bs{\alpha}},\overline{\fC}{}^{\bs{\alg{b}}},
\overline{\fC}{}^{\bs{\alg{c}}}\}|\psi^\kappa\rangle
=-3\sqrt{2}\overline{C}{}^{\bs{\alg{b}}}\overline{C}{}^{\bs{\alg{c}}}
(\gamma^{\bs{\alpha}})^\kappa{}_\lambda
|\psi^\lambda\rangle,\nonumber\\
&
\{\fR^{\bs{a}},\fF^{b\beta\alg{b}},\fF^{c\gamma\alg{c}}\}
|\psi^\kappa\rangle
=\frac{1}{2}(\gamma^{\bs{a}}\epsilon)^{bc}
\Bigl[\epsilon^{\beta\gamma}\delta^\kappa_\lambda
(\overline{\sC}\epsilon)^{\alg{b}\alg{c}}
+(\gamma^\kappa_\lambda\epsilon)^{\beta\gamma}
\epsilon^{\alg{b}\alg{c}}\Bigr]|\psi^\lambda\rangle,\nonumber\\
&
\{\fL^{\bs{\alpha}},\fF^{b\beta\alg{b}},\fF^{c\gamma\alg{c}}\}
|\psi^\kappa\rangle
=-\epsilon^{bc}
\Bigl[(\gamma^{\bs{\alpha}}\epsilon)^{\beta\gamma}
\delta^\kappa_\lambda
(\overline{\sC}\epsilon)^{\alg{b}\alg{c}}
-\frac{1}{\sqrt{2}}
\epsilon^{\beta\gamma}(\gamma^{\bs{\alpha}})^\kappa{}_\lambda
\epsilon^{\alg{b}\alg{c}}\Bigr]
|\psi^\lambda\rangle,\nonumber\\
&
\{\overline{\fC}{}^{\bs{\alg{a}}},
\fF^{b\beta\alg{b}},\fF^{c\gamma\alg{c}}\}|\psi^\kappa\rangle
=-\frac{3}{2}\overline{C}{}^{\bs{\alg{a}}}\epsilon^{bc}
\Bigl[2(\gamma^\kappa_\lambda\epsilon)^{\beta\gamma}
(\overline{\sC}\epsilon)^{\alg{b}\alg{c}}
+\epsilon^{\beta\gamma}\delta^\kappa_\lambda
\epsilon^{\alg{b}\alg{c}}\Bigr]
|\psi^\lambda\rangle.
\label{Bpsi}
\end{align}

$\bullet\;\{$fermionic$\}|\phi\rangle$
\begin{align}
&
\{\fR^{\bs{a}},\fR^{\bs{b}},\fF^{c\gamma\alg{c}}\}
|\phi^k\rangle
=-a^{\alg{c}}g^{\bs{a}\bs{b}}\epsilon^{kc}
|\psi^\gamma\rangle,\quad
\{\fR^{\bs{a}},\overline{\fC}{}^{\bs{\alg{b}}},\fF^{c\gamma\alg{c}}\}
|\phi^k\rangle
=\frac{3}{\sqrt{2}}
\overline{C}{}^{\bs{\alg{b}}}a^{\alg{c}}
(\gamma^{\bs{a}}\epsilon)^{kc}|\psi^\gamma\rangle,\nonumber\\
&
\{\fL^{\bs{\alpha}},\fL^{\bs{\beta}},\fF^{c\gamma\alg{c}}\}
|\phi^k\rangle
=-a^{\alg{c}}\epsilon^{kc}g^{\bs{\alpha}\bs{\beta}}
|\psi^\gamma\rangle,\quad
\{\fL^{\bs{\alpha}},\overline{\fC}{}^{\bs{\alg{b}}},
\fF^{c\gamma\alg{c}}\}|\phi^k\rangle
=\frac{3}{\sqrt{2}}
\overline{C}{}^{\bs{\alg{b}}}a^{\alg{c}}\epsilon^{kc}
(\gamma^{\bs{\alpha}})^\gamma{}_\lambda
|\psi^\lambda\rangle,\nonumber\\
&
\{\fR^{\bs{a}},\fL^{\bs{\beta}},\fF^{c\gamma\alg{c}}\}
|\phi^k\rangle
=-\frac{1}{2}a^{\alg{c}}(\gamma^{\bs{a}}\epsilon)^{kc}
(\gamma^{\bs{\beta}})^\gamma{}_\lambda|\psi^\lambda\rangle,\quad
\{\overline{\fC}{}^{\bs{\alg{a}}},\overline{\fC}{}^{\bs{\alg{b}}},
\fF^{c\gamma\alg{c}}\}|\phi^k\rangle
=-6\overline{C}{}^{\bs{\alg{a}}}\overline{C}{}^{\bs{\alg{b}}}
a^{\alg{c}}\epsilon^{kc}
|\psi^\gamma\rangle,\nonumber\\
&
\{\fF^{a\alpha\alg{a}},\fF^{b\beta\alg{b}},\fF^{c\gamma\alg{c}}\}
|\phi^k\rangle
=\bigl(b^\alg{a}a^\alg{b}a^\alg{c}+b^\alg{b}a^\alg{c}a^\alg{a}
+b^\alg{c}a^\alg{a}a^\alg{b}\bigr)\epsilon^{kl}
\bigl(\epsilon^{bc}\delta^a_l
\epsilon^{\gamma\alpha}\delta^\beta_\lambda
-\epsilon^{ca}\delta^b_l
\epsilon^{\beta\gamma}\delta^\alpha_\lambda\bigr)
|\psi^\lambda\rangle.
\label{Fphi}
\end{align}

$\bullet\;\{$fermionic$\}|\psi\rangle$
\begin{align}
&
\{\fR^{\bs{a}},\fR^{\bs{b}},\fF^{c\gamma\alg{c}}\}
|\psi^\kappa\rangle
=b^{\alg{c}}g^{\bs{a}\bs{b}}\epsilon^{\kappa\gamma}
|\phi^c\rangle,\quad
\{\fR^{\bs{a}},\overline{\fC}{}^{\bs{\alg{b}}},
\fF^{c\gamma\alg{c}}\}
|\psi^\kappa\rangle
=-\frac{3}{\sqrt{2}}\overline{C}{}^{\bs{\alg{b}}}b^{\alg{c}}
\epsilon^{\kappa\gamma}
(\gamma^{\bs{a}})^c{}_l|\phi^l\rangle,\nonumber\\
&
\{\fL^{\bs{\alpha}},\fL^{\bs{\beta}},\fF^{c\gamma\alg{c}}\}
|\psi^\kappa\rangle
=b^{\alg{c}}\epsilon^{\kappa\gamma}g^{\bs{\alpha}\bs{\beta}}
|\phi^c\rangle,\quad
\{\fL^{\bs{\alpha}},\overline{\fC}{}^{\bs{\alg{b}}},
\fF^{c\gamma\alg{c}}\}|\psi^\kappa\rangle
=-\frac{3}{\sqrt{2}}\overline{C}{}^{\bs{\alg{b}}}b^{\alg{c}}
(\gamma^{\bs{\alpha}}\epsilon)^{\kappa\gamma}
|\phi^c\rangle,\nonumber\\
&
\{\fR^{\bs{a}},\fL^{\bs{\beta}},\fF^{c\gamma\alg{c}}\}
|\psi^\kappa\rangle
=\frac{1}{2}b^{\alg{c}}(\gamma^{\bs{a}})^c{}_l
(\gamma^{\bs{\beta}}\epsilon)^{\kappa\gamma}
|\phi^l\rangle,\quad
\{\overline{\fC}{}^{\bs{\alg{a}}},\overline{\fC}{}^{\bs{\alg{b}}},
\fF^{c\gamma\alg{c}}\}|\psi^\kappa\rangle
=6\overline{C}{}^{\bs{\alg{a}}}\overline{C}{}^{\bs{\alg{b}}}b^{\alg{c}}
\epsilon^{\kappa\gamma}|\phi^c\rangle,\nonumber\\
&
\{\fF^{a\alpha\alg{a}},\fF^{b\beta\alg{b}},\fF^{c\gamma\alg{c}}\}
|\psi^\kappa\rangle
=\bigl(a^\alg{a}b^\alg{b}b^\alg{c}+a^\alg{b}b^\alg{c}b^\alg{a}
+a^\alg{c}b^\alg{a}b^\alg{b}\bigr)\epsilon^{\kappa\lambda}
\bigl(\epsilon^{bc}\delta^a_l
\epsilon^{\gamma\alpha}\delta^\beta_\lambda
-\epsilon^{ca}\delta^b_l
\epsilon^{\beta\gamma}\delta^\alpha_\lambda\bigr)
|\phi^l\rangle.
\label{Fpsi}
\end{align}

Note that the combination $\epsilon^{bc}\delta^a_l
\epsilon^{\gamma\alpha}\delta^\beta_\lambda
-\epsilon^{ca}\delta^b_l
\epsilon^{\beta\gamma}\delta^\alpha_\lambda$ in the right-hand-side of 
the last lines in \eqref{Fphi} and \eqref{Fpsi} does not seem to
respect the totally symmetric property of the left-hand-side at the
first sight.
However, there is an interesting cyclic property in this combination,
\begin{align}
\epsilon^{bc}\delta^a_l
\epsilon^{\gamma\alpha}\delta^\beta_\lambda
-\epsilon^{ca}\delta^b_l
\epsilon^{\beta\gamma}\delta^\alpha_\lambda
=\epsilon^{ca}\delta^b_l
\epsilon^{\alpha\beta}\delta^\gamma_\lambda
-\epsilon^{ab}\delta^c_l
\epsilon^{\gamma\alpha}\delta^\beta_\lambda
=\epsilon^{ab}\delta^c_l
\epsilon^{\beta\gamma}\delta^\alpha_\lambda
-\epsilon^{bc}\delta^a_l
\epsilon^{\alpha\beta}\delta^\gamma_\lambda,
\end{align}
which can be easily proved from
$\epsilon^{bc}\delta^a_l+\epsilon^{ca}\delta^b_l
+\epsilon^{ab}\delta^c_l=0$ 
and 
$\epsilon^{\beta\gamma}\delta^\alpha_\lambda
+\epsilon^{\gamma\alpha}\delta^\beta_\lambda
+\epsilon^{\alpha\beta}\delta^\gamma_\lambda=0$.

\section{Formulas for exceptional superalgebra}
In this appendix we shall see how the expression of the totally
symmetrized generators of $\alg{psu}(2|2)\ltimes{\mathbb R}^3$ in the
previous appendix is lifted to the case of the exceptional
superalgebra $\alg{d}(2,1;\varepsilon)$.
Our strategy is to search for a product of the generators which looks
simpler but has the same action as the totally symmetrized generators.
A naive guess is to rewrite the gamma matrices $\gamma^k{}_l$ and
$\gamma^\kappa{}_\lambda$ in \eqref{Bphi}-\eqref{Fpsi} into generators
$\fR$ and $\fL$, and the coefficients $a$, $b$ and $C$ into generators
$\fF$ and $\fC$.
Since the the generators do not commute with each other in general, we 
have to take care of the ordering carefully.
Here for simplicity we shall abbreviate the indices of the states as
$|\phi\rangle=|\phi^k_m\rangle$, $|\psi\rangle=|\psi^\kappa_r\rangle$,
because the formulas do not depend on the indices.

$\bullet\;\{$bosonic$\}|\phi\rangle$
\begin{align}
&
\{\fR^{\bs{a}},\fR^{\bs{b}},\fR^{\bs{c}}\}|\phi\rangle
=(g^{\bs{b}\bs{c}}\fR^{\bs{a}}
+g^{\bs{c}\bs{a}}\fR^{\bs{b}}
+g^{\bs{a}\bs{b}}\fR^{\bs{c}})
|\phi\rangle,\nonumber\\
&
\{\fR^{\bs{a}},\fR^{\bs{b}},
\fC{}^{\bs{\alg{c}}}\}|\phi\rangle
=3\fC^{\bs{\alg{c}}}g^{\bs{a}\bs{b}}|\phi\rangle,\quad
\{\fR^{\bs{a}},\fC{}^{\bs{\alg{b}}},\fC{}^{\bs{\alg{c}}}\}
|\phi\rangle
=3\fR^{\bs{a}}\{\fC{}^{\bs{\alg{b}}},\fC{}^{\bs{\alg{c}}}\}
|\phi\rangle,\nonumber\\
&
\{\fR^{\bs{a}},\fF^{b\beta\alg{b}},\fF^{c\gamma\alg{c}}\}
|\phi\rangle
=-\epsilon^{\beta\gamma}
\Bigl[\alpha\epsilon^{bc}\epsilon^{\alg{b}\alg{c}}\fR^{\bs{a}}
+\gamma(\gamma^{\bs{a}}\epsilon)^{bc}
(\sfC\epsilon)^{\alg{b}\alg{c}}\Bigr]
|\phi\rangle,\nonumber\\
&
\{\fL^{\bs{\alpha}},\fF^{b\beta\alg{b}},\fF^{c\gamma\alg{c}}\}
|\phi\rangle
=\frac{1}{2}(\gamma^{\bs{\alpha}}\epsilon)^{\beta\gamma}
\Bigl[\alpha(\sfR\epsilon)^{bc}\epsilon^{\alg{b}\alg{c}}
+\gamma\epsilon^{bc}(\sfC\epsilon)^{\alg{b}\alg{c}}\Bigr]
|\phi\rangle,\nonumber\\
&
\{\fC{}^{\bs{\alg{a}}},\fF^{b\beta\alg{b}},\fF^{c\gamma\alg{c}}\}
|\phi\rangle
=\frac{-1}{2}\epsilon^{\beta\gamma}
\Bigl[(\sfR\epsilon)^{bc}
\bigl(3\gamma\{\fC^{\alg{a}},(\sfC\epsilon)^{\alg{bc}}\}
-\alpha(\gamma^{\bs{\alg{a}}}\epsilon)^{\alg{bc}}\bigr)
+(3\alpha-\gamma)\epsilon^{bc}\epsilon^{\alg{bc}}\fC^{\bs{\alg{a}}}
\Bigr]|\phi\rangle.
\end{align}

$\bullet\;\{$bosonic$\}|\psi\rangle$
\begin{align}
&
\{\fL^{\bs{\alpha}},\fL^{\bs{\beta}},\fL^{\bs{\gamma}}\}
|\psi\rangle
=(g^{\bs{\beta}\bs{\gamma}}\fL^{\bs{\alpha}}
+g^{\bs{\gamma}\bs{\alpha}}\fL^{\bs{\beta}}
+g^{\bs{\alpha}\bs{\beta}}\fL^{\bs{\gamma}})
|\psi\rangle,\nonumber\\
&
\{\fL^{\bs{\alpha}},\fL^{\bs{\beta}},\fC^{\bs{\alg{c}}}\}
|\psi\rangle
=3\fC^{\bs{\alg{c}}}g^{\bs{\alpha}\bs{\beta}}|\psi\rangle,\quad
\{\fL^{\bs{\alpha}},\fC^{\bs{\alg{b}}},\fC^{\bs{\alg{c}}}\}
|\psi\rangle
=3\fL^{\bs{\alpha}}\{\fC^{\bs{\alg{b}}},\fC^{\bs{\alg{c}}}\}
|\psi\rangle,\nonumber\\
&
\{\fR^{\bs{a}},\fF^{b\beta\alg{b}},\fF^{c\gamma\alg{c}}\}
|\psi\rangle
=\frac{1}{2}(\gamma^{\bs{a}}\epsilon)^{bc}
\Bigl[\beta(\sfL\epsilon)^{\beta\gamma}\epsilon^{\alg{bc}}
+\gamma\epsilon^{\beta\gamma}(\sfC\epsilon)^{\alg{bc}}\Bigr]
|\psi\rangle,\nonumber\\
&
\{\fL^{\bs{\alpha}},\fF^{b\beta\alg{b}},\fF^{c\gamma\alg{c}}\}
|\psi\rangle
=-\epsilon^{bc}
\Bigl[\beta\epsilon^{\beta\gamma}\epsilon^{\alg{bc}}\fL^{\bs{\alpha}}
+\gamma(\gamma^{\bs{\alpha}}\epsilon)^{\beta\gamma}
(\sfC\epsilon)^{\alg{bc}}\Bigr]
|\psi\rangle,\nonumber\\
&
\{\fC^{\bs{\alg{a}}},\fF^{b\beta\alg{b}},\fF^{c\gamma\alg{c}}\}
|\psi\rangle
=\frac{-1}{2}\epsilon^{bc}
\Bigl[(\sfL\epsilon)^{\beta\gamma}
\bigl(3\gamma\{\fC^{\alg{a}},(\sfC\epsilon)^{\alg{bc}}\}
-\beta(\gamma^{\bs{\alg{a}}}\epsilon)^{\alg{bc}}\bigr)
+(3\beta-\gamma)\epsilon^{\beta\gamma}\epsilon^{\alg{bc}}
\fC^{\bs{\alg{a}}}\Bigr]|\psi\rangle.
\end{align}

$\bullet\;\{$fermionic$\}|\phi\rangle$
\begin{align}
&
\{\fR^{\bs{a}},\fR^{\bs{b}},\fF^{c\gamma\alg{c}}\}
|\phi\rangle
=g^{\bs{a}\bs{b}}\fF^{c\gamma\alg{c}}
|\phi\rangle,\quad
\{\fR^{\bs{a}},\fC^{\bs{\alg{b}}},\fF^{c\gamma\alg{c}}\}
|\phi\rangle
=\frac{1}{\sqrt{2}}(\gamma^{\bs{a}})^c{}_d
\Bigl[2\fF^{d\gamma\alg{c}}\fC^{\bs{\alg{b}}}
+\fC^{\bs{\alg{b}}}\fF^{d\gamma\alg{c}}\Bigr]
|\phi\rangle,\nonumber\\
&
\{\fL^{\bs{\alpha}},\fL^{\bs{\beta}},\fF^{c\gamma\alg{c}}\}
|\phi\rangle
=g^{\bs{\alpha}\bs{\beta}}\fF^{c\gamma\alg{c}}
|\phi\rangle,\quad
\{\fL^{\bs{\alpha}},\fC^{\bs{\alg{b}}},\fF^{c\gamma\alg{c}}\}
|\phi\rangle
=\frac{-1}{\sqrt{2}}(\gamma^{\bs{\alpha}})^\gamma{}_\delta
\Bigr[2\fC^{\bs{\alg{b}}}\fF^{c\delta\alg{c}}
+\fF^{c\delta\alg{c}}\fC^{\bs{\alg{b}}}\Bigr]
|\phi\rangle,\nonumber\\
&
\{\fR^{\bs{a}},\fL^{\bs{\beta}},\fF^{c\gamma\alg{c}}\}
|\phi\rangle
=\frac{-1}{2}(\gamma^{\bs{a}})^c{}_d
(\gamma^{\bs{\beta}})^\gamma{}_\delta\fF^{d\delta\alg{c}}
|\phi\rangle,\nonumber\\
&
\{\fC^{\bs{\alg{a}}},\fC^{\bs{\alg{b}}},\fF^{c\gamma\alg{c}}\}
|\phi\rangle
=\frac{1}{2}
\Bigl[3\{\{\fC^{\bs{\alg{a}}},\fC^{\bs{\alg{b}}}\},\fF^{c\gamma\alg{c}}\}
-g^{\bs{\alg{a}}\bs{\alg{b}}}\fF^{c\gamma\alg{c}}\Bigr]
|\phi\rangle,\nonumber\\
&
\{\fF^{a\alpha\alg{a}},\fF^{b\beta\alg{b}},\fF^{c\gamma\alg{c}}\}
|\phi\rangle
=\frac{-1}{2}\Bigl\{3\gamma\Bigl[
\fF^{c\gamma\alg{c}}
(\sfR\epsilon)^{ab}\epsilon^{\alpha\beta}(\sfC\epsilon)^{\alg{a}\alg{b}}
+\epsilon^{ab}(\sfL\epsilon)^{\alpha\beta}(\sfC\epsilon)^{\alg{a}\alg{b}}
\fF^{c\gamma\alg{c}}
-\frac{1}{2}\epsilon^{ab}\epsilon^{\alpha\beta}\epsilon^{\alg{a}\alg{b}}
\fF^{c\gamma\alg{c}}\Bigr]\nonumber\\
&\quad+\Bigl[
\alpha\delta^c_d\epsilon^{ab}
(\gamma^\gamma_\delta\epsilon)^{\alpha\beta}
(\gamma^\alg{c}_\alg{d}\epsilon)^{\alg{a}\alg{b}}
+\beta(\gamma^c_d\epsilon)^{ab}
\delta^\gamma_\delta\epsilon^{\alpha\beta}
(\gamma^\alg{c}_\alg{d}\epsilon)^{\alg{a}\alg{b}}
+\gamma(\gamma^c_d\epsilon)^{ab}
(\gamma^\gamma_\delta\epsilon)^{\alpha\beta}
\delta^\alg{c}_\alg{d}\epsilon^{\alg{a}\alg{b}}\Bigr]
\fF^{d\delta\alg{d}}\Bigr\}|\phi\rangle.
\end{align}

$\bullet\;\{$fermionic$\}|\psi\rangle$
\begin{align}
&
\{\fR^{\bs{a}},\fR^{\bs{b}},\fF^{c\gamma\alg{c}}\}
|\psi\rangle
=g^{\bs{a}\bs{b}}\fF^{c\gamma\alg{c}}
|\psi\rangle,\quad
\{\fR^{\bs{a}},\fC^{\bs{\alg{b}}},\fF^{c\gamma\alg{c}}\}
|\psi\rangle
=\frac{-1}{\sqrt{2}}(\gamma^{\bs{a}})^c{}_d
\Bigl[2\fC^{\bs{\alg{b}}}\fF^{d\gamma\alg{c}}
+\fF^{d\gamma\alg{c}}\fC^{\bs{\alg{b}}}\Bigr]
|\psi\rangle,\nonumber\\
&
\{\fL^{\bs{\alpha}},\fL^{\bs{\beta}},\fF^{c\gamma\alg{c}}\}
|\psi\rangle
=g^{\bs{\alpha}\bs{\beta}}\fF^{c\gamma\alg{c}}
|\psi\rangle,\quad
\{\fL^{\bs{\alpha}},\fC^{\bs{\alg{b}}},\fF^{c\gamma\alg{c}}\}
|\psi\rangle
=\frac{1}{\sqrt{2}}(\gamma^{\bs{\alpha}})^\gamma{}_\delta
\Bigr[2\fF^{c\delta\alg{c}}\fC^{\bs{\alg{b}}}
+\fC^{\bs{\alg{b}}}\fF^{c\delta\alg{c}}\Bigr]
|\psi\rangle,\nonumber\\
&
\{\fR^{\bs{a}},\fL^{\bs{\beta}},\fF^{c\gamma\alg{c}}\}
|\psi\rangle
=\frac{-1}{2}(\gamma^{\bs{a}})^c{}_d
(\gamma^{\bs{\beta}})^\gamma{}_\delta\fF^{d\delta\alg{c}}
|\psi\rangle,\nonumber\\
&
\{\fC^{\bs{\alg{a}}},\fC^{\bs{\alg{b}}},\fF^{c\gamma\alg{c}}\}
|\psi\rangle
=\frac{1}{2}
\Bigl[3\{\{\fC^{\bs{\alg{a}}},\fC^{\bs{\alg{b}}}\},\fF^{c\gamma\alg{c}}\}
-g^{\bs{\alg{a}}\bs{\alg{b}}}\fF^{c\gamma\alg{c}}\Bigr]
|\psi\rangle,\nonumber\\
&
\{\fF^{a\alpha\alg{a}},\fF^{b\beta\alg{b}},\fF^{c\gamma\alg{c}}\}
|\psi\rangle
=\frac{-1}{2}\Bigl\{3\gamma\Bigl[
(\sfR\epsilon)^{ab}\epsilon^{\alpha\beta}(\sfC\epsilon)^{\alg{a}\alg{b}}
\fF^{c\gamma\alg{c}}
+\fF^{c\gamma\alg{c}}
\epsilon^{ab}(\sfL\epsilon)^{\alpha\beta}(\sfC\epsilon)^{\alg{a}\alg{b}}
-\frac{1}{2}\epsilon^{ab}\epsilon^{\alpha\beta}\epsilon^{\alg{a}\alg{b}}
\fF^{c\gamma\alg{c}}\Bigr]\nonumber\\
&\quad+\Bigl[
\alpha\delta^c_d\epsilon^{ab}
(\gamma^\gamma_\delta\epsilon)^{\alpha\beta}
(\gamma^\alg{c}_\alg{d}\epsilon)^{\alg{a}\alg{b}}
+\beta(\gamma^c_d\epsilon)^{ab}
\delta^\gamma_\delta\epsilon^{\alpha\beta}
(\gamma^\alg{c}_\alg{d}\epsilon)^{\alg{a}\alg{b}}
+\gamma(\gamma^c_d\epsilon)^{ab}
(\gamma^\gamma_\delta\epsilon)^{\alpha\beta}
\delta^\alg{c}_\alg{d}\epsilon^{\alg{a}\alg{b}}\Bigr]
\fF^{d\delta\alg{d}}\Bigr\}|\psi\rangle.
\end{align}

In deriving all these expressions the formulas we need are
\begin{align}
&\{\fJ^A,\fJ^B,\fJ^C\}=\frac{3}{2}\{\{\fJ^A,\fJ^B\},\fJ^C\}
+\frac{1}{2}(-)^{BC}
\Bigl([\fJ^A,[\fJ^C,\fJ^B]]+[[\fJ^A,\fJ^C],\fJ^B]\Bigr),
\nonumber\\
&\{\fR^{\bs{a}},\fR^{\bs{b}}\}|\phi\rangle
=g^{\bs{a}\bs{b}}|\phi\rangle,\quad
\{\fL^{\bs{\alpha}},\fL^{\bs{\beta}}\}|\psi\rangle
=g^{\bs{\alpha}\bs{\beta}}|\psi\rangle,\nonumber\\
&\{\fF^{b\beta\alg{b}},\fF^{c\gamma\alg{c}}\}|\phi\rangle
=-\epsilon^{\beta\gamma}\bigl[
\gamma(\sfR\epsilon)^{bc}(\sfC\epsilon)^{\alg{b}\alg{c}}
+\frac{\alpha}{2}\epsilon^{bc}\epsilon^{\alg{b}\alg{c}}
\bigr]|\phi\rangle,
\nonumber\\
&
\{\fF^{b\beta\alg{b}},\fF^{c\gamma\alg{c}}\}|\psi\rangle
=-\epsilon^{bc}\bigl[
\gamma(\sfL\epsilon)^{\beta\gamma}(\sfC\epsilon)^{\alg{b}\alg{c}}
+\frac{\beta}{2}\epsilon^{\beta\gamma}\epsilon^{\alg{b}\alg{c}}
\bigr]|\psi\rangle,
\end{align}
while the following formulas
\begin{align}
&
\fC^{\bs{\alg{a}}}\fC_{\bs{\alg{a}}}|\phi\rangle
=\frac{-1}{2\gamma^2}(2\alpha\beta+\alpha^2)|\phi\rangle,\quad
\fC^{\bs{\alg{a}}}\fC_{\bs{\alg{a}}}|\psi\rangle
=\frac{-1}{2\gamma^2}(2\alpha\beta+\beta^2)|\psi\rangle,
\nonumber\\
&
\fF^{a\alpha\alg{a}}(\sfC)^\alg{b}{}_\alg{a}|\phi\rangle
=-\frac{\alpha+2\beta}{\sqrt{2}\gamma}
\fF^{a\alpha\alg{b}}|\phi\rangle,
\quad
(\sfC)^\alg{b}{}_\alg{a}\fF^{a\alpha\alg{a}}|\phi\rangle
=\frac{2\alpha+\beta}{\sqrt{2}\gamma}
\fF^{a\alpha\alg{b}}|\phi\rangle,
\nonumber\\
&\fF^{a\alpha\alg{a}}(\sfC)^\alg{b}{}_\alg{a}|\psi\rangle
=-\frac{2\alpha+\beta}{\sqrt{2}\gamma}
\fF^{a\alpha\alg{b}}|\psi\rangle,
\quad
(\sfC)^\alg{b}{}_\alg{a}\fF^{a\alpha\alg{a}}|\psi\rangle
=\frac{\alpha+2\beta}{\sqrt{2}\gamma}
\fF^{a\alpha\alg{b}}|\psi\rangle,
\end{align}
are also useful for further calculations in proving the compatibility.

\end{document}